\documentclass[10.5pt,compsoc]{tst}
\usepackage{graphicx}
\providecommand{\IEEEPARstart}[2]{\textbf{#1}#2}
\usepackage{footmisc}
\usepackage{subfigure}
\usepackage{url}
\usepackage{multirow}
\usepackage[noadjust]{cite}
\usepackage{amsmath,amsthm}
\usepackage{amssymb,amsfonts}
\usepackage{booktabs}
\usepackage{color}
\usepackage{ccaption}
\usepackage{booktabs}
\usepackage{float}
\usepackage{fancyhdr}
\usepackage{caption}
\usepackage{xcolor,stfloats}
\usepackage{comment}
\graphicspath{{figures/}{../jpeg/}}
\usepackage{cuted}  
\usepackage{captionhack}
\usepackage{epstopdf}

\usepackage{cite}
\usepackage{amsmath,amssymb,amsfonts}
\usepackage{graphicx}
\usepackage{textcomp}
\usepackage{xcolor}
\usepackage{ifpdf}
\usepackage{subfigure}
\makeatletter
\newif\if@restonecol
\makeatother
\usepackage[ruled,vlined]{algorithm2e}
\usepackage{amssymb}

\headevenname{\normalsize{\textbf{\emph{Tsinghua Science and Technology, xxxxx}} 20xx, xx(x): xxx--xxx}}%
\headoddname{{\sf Chaoyue Zhang et al.:}\quad {\textbf{\emph{Resource Management and Trajectory Optimization for UAV-IRS Assisted Maritime Edge Computing Networks}}}}%

\newtheoremstyle{mystyle}{0pt}{0pt}{\normalfont}{1em}{\bf}{}{1em}{}
\theoremstyle{mystyle}
\renewcommand\figurename{Fig.~}

\newcommand{\upcite}[1]{\textsuperscript{\cite{#1}}}

\newcommand{\nop}[1]{}

\makeatletter
\renewcommand{\@biblabel}[1]{[#1]\hfill}
\makeatother

\begin{document}

\thispagestyle{empty}

\clearpage

\hyphenpenalty=50000

\makeatletter
\newcommand\mysmall{\@setfontsize\mysmall{7}{9.5}}

\newenvironment{tablehere}
  {\def\@captype{table}}
  {}
\newenvironment{figurehere}
  {\def\@captype{figure}}
  {}

\thispagestyle{plain}%
\thispagestyle{empty}%

\let\temp\footnote
\renewcommand \footnote[1]{\temp{\normalsize #1}}
{}
\vspace*{-40pt}
\noindent{\normalsize\textbf{\scalebox{0.885}[1.0]{\makebox[5.9cm][s]
{TSINGHUA\, SCIENCE\, AND\, TECHNOLOGY}}}}

\vskip .2mm
{\normalsize
\textbf{
\hspace{-5mm}
\scalebox{1}[1.0]{\makebox[5.6cm][s]{%
I\hfill S\hfill S\hfill N\hfill{\color{white}%
l\hfill l\hfill}1\hfill0\hfill0\hfill7\hfill-\hfill0\hfill2\hfill1\hfill4
\hfill \color{white}{\quad 0\hfill ?\hfill /\hfill ?\hfill ?\quad p\hfill p\hfill  ?\hfill ?\hfill ?\hfill --\hfill ?\hfill ?\hfill ?}\hfill}}}}

\vskip .2mm
{\normalsize
\textbf{
\hspace{-5mm}
\scalebox{1}[1.0]{\makebox[5.6cm][s]{%
DOI:~\hfill~\hfill1\hfill0\hfill.\hfill2\hfill6\hfill5\hfill9\hfill9\hfill/\hfill T\hfill S\hfill T\hfill.\hfill2\hfill0\hfill 2\hfill 4\hfill.\hfill9\hfill0\hfill1\hfill0\hfill 0\hfill 7\hfill 4}}}}

\vskip .2mm\noindent
{\normalsize\textbf{\scalebox{1}[1.0]{\makebox[5.6cm][s]{%
\color{black}{V\hfill o\hfill l\hfill u\hfill m\hfill%
e\hspace{0.356em}xx,\hspace{0.356em}N\hfill u\hfill%
m\hfill b\hfill e\hfill r\hspace{0.356em}x,\hspace{0.356em}%
x\hfill x\hfill x\hfill x\hfill x\hfill%
x\hfill x\hfill \hspace{0.356em}2\hfill0\hfill x\hfill x}}}}}\\

\begin{strip}
{\center
{\LARGE\textbf{
 Resource Management and Trajectory Optimization for UAV-IRS Assisted Maritime Edge Computing Networks}}
\vskip 9mm}

{\center {\sf \large
Chaoyue Zhang, Bin Lin*, Xu Hu, Shuang Qi, Liping Qian, Yuan Wu
}
\vskip 5mm}

\centering{
\begin{tabular}{p{160mm}}

{\normalsize
\linespread{1.6667} %
\noindent
\bf{Abstract:} {\sf
With the exponential growth of maritime wireless devices and the rapid development of maritime applications, traditional maritime communication networks encounter communication and computation limitations in supporting computation-intensive and latency-critical tasks. Edge computing and Intelligent Reflecting Surface (IRS) have emerged as promising techniques to improve communication and computation services for maritime devices with limited computation capabilities and battery capacity. This paper studies an IRS Mounted on Unmanned Aerial Vehicle (UIRS) assisted maritime edge computing network, in which the UIRS is deployed to assist the transmission from Unmanned Surface Vehicles (USVs) to the edge server via Non-Orthogonal Multiple Access (NOMA) protocol. We propose a resource management and trajectory optimization scheme by jointly optimizing subslot duration, offloading ratios, transmit power, edge computation capability allocation, UIRS phase shifts and UIRS trajectory, aiming at minimizing the overall energy consumption. Since the non-convex nature of the optimization problem, we propose a two-layered method by decomposing the original problem into two subproblems. The top-layered subproblem is solved by the Semi-Definite Relaxation (SDR) method and the underlying-layered subproblem is solved by the Deep Deterministic Policy Gradient (DDPG) algorithm. Numerical results demonstrate that our proposed scheme can effectively and efficiently reduce overall energy consumption.
}
}

\end{tabular}
}
\vskip 6mm

\vskip -3mm
\small\end{strip}

\thispagestyle{plain}%
\thispagestyle{empty}%
\makeatother
\pagestyle{tstheadings}

\begin{figure}[b]
\vskip -6mm
\begin{tabular}{p{44mm}}
\toprule\\
\end{tabular}
\vskip -5.0mm
\noindent
\setlength{\tabcolsep}{1pt}
\begin{tabular}{p{1.5mm}p{79.5mm}}

$\bullet$& Chaoyue Zhang, Bin Lin, Xu Hu, and Shuang Qi are with the Information Science and Technology College, Dalian Maritime University, Dalian 116026, China. E-mail: zcy\_11335577@163.com; binlin@dlmu.edu.cn; huxu@dlmu.edu.cn; qishuang\_0315@163.com.\\ 
$\bullet$& Liping Qian is with the College of Information Engineering, Zhejiang University of Technology, Hangzhou 310014, China. E-mail: lpqian@zjut.edu.cn. \\
$\bullet$& Yuan Wu is with the State Key Laboratory of Internet of Things for Smart City, University of Macau, Macau, China E-mail: yuanwu@um.edu.mo. \\
$\sf{*}$&
To whom correspondence should be addressed. \\
          &          Manuscript received: 2023-11-16; revised: 2024-01-16; accepted: 2024-04-13

\end{tabular}
\end{figure}\large

\large

\vspace{3.5mm}

\section{Introduction}

\IEEEPARstart Currently, a large amount of maritime smart facilities, such as Unmanned Surface Vehicles (USVs) and buoys, have attracted attention in both academia and maritime industry. These facilities have the potential to support a burgeoning array of maritime applications, such as intelligent navigation, real-time video analysis and emergency rescue\upcite{r2,r3}, leading to the generation of significant data traffic, measured in megabits. The rapid growth in data traffic is driving ever-growing demands for communication and computation resources. Due to the limited computation capability and energy capacity, these smart facilities encounter difficulties in efficiently processing extensive maritime tasks and communicating with shore base stations. For these, the maritime edge computing network with high communication and computation capabilities is envisioned as a promising solution to compute the intensive maritime data traffic by deploying edge servers co-located with the infrastructures\upcite{r54,r55,r56}. It can offload computation tasks generated by various maritime smart facilities to edge servers to reduce the transmission delay and computation energy consumption of facilities. However, the quality of the computation offloading may not satisfy the maritime data traffic requirements. This limitation stems from the signal transmission link from maritime smart facilities and edge servers always suffering from signal attenuation and limited spectrum resources, resulting in a low data rate\upcite{r51,r12,r52}. Therefore, it is critical to improve the computation offloading performance of maritime edge computing networks. 

Intelligent Reflecting Surface (IRS) has recently emerged as a novel technology to realize intelligent and reconfigurable wireless communication environments\upcite{r41,r42}. It can achieve intelligent control of the physical channel by adjusting the amplitude and phase shift of the passive reflective elements in a programmable manner of the IRS controller. As a result, the passive beamforming reflection for signals can be realized in a full-duplex mode without additional energy consumption, self-interference and noise amplification\upcite{r12,r13}. Integrating IRS and edge computing is expected to develop maritime wireless networks to provide communication and computation services. By deploying IRS between maritime facilities and edge servers, IRS provides auxiliary links for maritime facilities through passive beamforming, thus the computation task data can be offloaded to edge servers with low computation delay and high data rate in maritime edge computing networks\upcite{r38}. However, finding a stable area for deploying the IRS device is a challenging task owing to the floating sea surface. Due to the advantages of flexibility and high probability Line-of-Sight (LoS) of Unmanned Aerial Vehicle (UAV), IRS mounted on UAV (UIRS) provides a feasible solution to support maritime edge computing networks\upcite{r36}.

This paper focuses on the energy consumption of processing tasks in UIRS-assisted maritime edge computing networks. Two important decisions including UIRS trajectory and resource management are required to be investigated in this paper. Specifically, the position of UIRS mainly determines the quality of reconfigurable wireless propagation environments and the channel capacity, which can significantly influence the transmission performance of maritime edge computing networks. For instance, Mu et al.\upcite{r13} studied the performance gain achieved with IRS can be significantly improved by optimizing the deployment location based on numerical simulations. To achieve the proper UIRS trajectory optimization, it is critical to balance the flight duration and hovering duration. Note that the flying duration is used to reach specified positions of UIRS and the hovering duration is used to transmit the task data to the edge server when the UIRS is in a hovering state. In this paper, over-allocation of the flying duration can increase unnecessary flying energy consumption and also reduce the size of offloading data, resulting in large local computation energy consumption. Conversely, offloading excessive task data to the edge server could require high transmit power to support the required large transmission rate, resulting in large transmission energy consumption of maritime facilities. Therefore, it is necessary to consider both the dynamic offloading task data, communication and computation resources in making UIRS trajectory and resource management decisions. Given the highly dynamic and unpredictable characteristics of these factors, it is challenging to make resource management and UIRS trajectory decisions, while striving to achieve low energy consumption for processing tasks in UIRS-assisted maritime edge computing networks.

In this paper, we propose a Resource Management and Trajectory Optimization (RMTO) scheme in UIRS-assisted maritime edge computing networks. Our objective is to minimize the overall energy consumption by jointly optimizing the subslot duration, offloading ratios, transmit power, computation capability allocation, UIRS phase shifts and UIRS trajectory. Since the formulated overall energy consumption minimization problem is non-convex, we decompose it into two subproblems, solving in an alternating optimization approach to attain the near optimal solution. Extensive numerical results are presented to evaluate the convergence of the proposed scheme. Compared with baseline schemes, the proposed scheme can effectively reduce the overall energy consumption in different simulation environments. The main contributions of this paper are summarized as follows.  

$\bullet$ We investigate a UIRS-assisted maritime edge computing network, where the UIRS provides auxiliary links for USVs and intelligently controls the received signal to improve the transmission performance between USVs and the edge server. Each USV has computation task data that can be computed locally or partially offloaded to the edge server. A joint optimization problem is proposed to minimize the overall energy consumption for processing computation tasks of USVs, with the constraints of computation capabilities and offloading data requirements of USVs.

$\bullet$ To solve the non-convex optimization problem, we decompose the original problem into two tractable subproblems. Specifically, in the top-layered subproblem, we derive the closed-form expression of the transmit power and then apply the Semi-Definite Relaxation (SDR) method to optimize the UIRS phase shifts. To solve the underlying-layered subproblem, we transform it as a Markov Decision Process (MDP). Due to the huge state space and continuous action in the problem, we employ a deep neural network to characterize the state space and action space. Specifically, we design an algorithm based on Deep Deterministic Policy Gradient (DDPG) to obtain the optimal solutions for the subslot duration, offloading ratios, computation capability allocation and UIRS trajectory.

The remainder of this paper is organized as follows. Section 2 reviews the related works. Section 3 illustrates the system model and problem formulation for UIRS-assisted maritime edge computing networks. Section 4 decomposes the original problem into two subproblems and proposes algorithms to solve them. Section 5 presents numerical results to evaluate the performance of the proposed scheme. Finally, Section 6 concludes this paper.

\textit{Notations}: $|x|$ denotes the absolute value of the number $x$. $\|\mathbf{x}\|$ denotes the Euclidean norm of vector $\mathbf{x}$. $\mathbf{x}^{\rm T}$ denotes the transpose matrix of vector $\mathbf{x}$. $\mathbf{x}^{\rm H}$ denotes the conjugate transpose matrix of vector $\mathbf{x}$. $diag(\mathbf{x})$ denotes the diagonal matrix of vector $\mathbf{x}$. For a matrix $\mathbf{X}$. $rank(\mathbf{X})$ denotes the rank of matrix $\mathbf{X}$. $tr(\mathbf{X})$ denotes the trace of matrix $\mathbf{X}$. $\mathbf{X} \succeq 0$ denotes $\mathbf{X}$ as a semi-definite matrix. $\mathbf{X}\otimes\mathbf{Y}$ denotes the Kronecker product of two matrices $\mathbf{X}$ and $\mathbf{Y}$. $\mathbb{C}^{M \times N}$ denotes the space of $M\times N$ complex matrix. $\mathbb{E}[\cdot]$ denotes the expectation.

\section{Related Works}
\subsection{Fixed IRS-Assisted Edge Networks}
Recently, the fixed IRS is expected to improve the communication performance of edge networks\upcite{r38,r17,r18,r49,r50,r53}. Wang et al.\upcite{r38} optimized user association, computation offloading, resource allocation and IRS phase shifts to minimize the long-term energy consumption in an edge heterogeneous network. Yang et al.\upcite{r17} considered an IRS-aided multi-device Mobile Edge Computing (MEC) system, with the objective of minimizing the overall energy consumption by jointly optimizing binary offloading modes, CPU frequencies, offloading power, offloading times and IRS phase shifts. In order to greatly improve the spectral efficiency and support massive connectivity, Xu et al.\upcite{r18} investigated an IRS-assisted Non-Orthogonal Multiple Access (NOMA)-based MEC system, aiming at minimizing the energy consumption by jointly optimizing transmit power, offloading partitions, transmission time and IRS phase shifts. Zhu et al.\upcite{r49} studied the relationship between computation resource allocation and offloading strategy at an intersection to minimize the computation throughput in the IRS-aided MEC-served vehicular networks. Li et al.\upcite{r50} investigated an IRS-aided single-cell multi-user MEC system where an IRS is deployed to support the communication between a Base Station (BS) and users, and optimized IRS phase shifts, transmission data size, transmission rate, transmit powers, transmission time and the decoding order, with the objective of minimizing the energy consumption. Chen et al.\upcite{r53} developed a dynamic IRS beamforming framework to enhance the sum computation rate of an IRS-aided MEC system, and optimized the offloading modes, time allocation, transmit powers and IRS beamforming design. The above-mentioned research on IRS-assisted edge networks based on fixed IRS has achieved some performance improvements. However, from the deployment perspective, the appropriate position for IRS installation is usually difficult in ocean areas. Therefore, this paper considers the UIRS to assist the computation offloading in maritime edge computing networks.

\subsection{UAV-IRS-Assisted Edge Networks}
The position of UAV-IRS can be flexibly optimized to further improve the performance of UAV-IRS-assisted edge networks, thereby having drawn increasing attention\upcite{r11,r25,r15,r14,r10,r24}. Shnaiwer et al.\upcite{r11} studied a joint IRS phase shifts and offloading decisions problem in a UAV-assisted MEC system, aiming at minimizing the overall energy consumption. Shafique et al.\upcite{r25} considered UAV-only, IRS-only and integrated UAV-IRS transmission modes in a UAV-IRS relaying system, and used the derived expressions to optimize the UAV altitude and the number of IRS reflection elements in different modes. Compared to the static deployment of UIRS, the UIRS trajectory optimization allows for adaptive adjustments based on channel conditions and quality-of-service requirements, providing a new Degree-of-Freedom (DoF) for network performance optimization\upcite{r15}. Zhai et al.\upcite{r14} considered a UIRS-assisted MEC system and formulated a joint UAV trajectory, IRS passive beamforming and resource allocation problem to maximize the energy efficiency. Ai et al.\upcite{r10} formulated an energy consumption minimization problem by jointly optimizing offloading decisions of USVs, transmit power of USVs, IRS phase shifts and UAV trajectory in IRS-aided wireless inland ship MEC networks. Su et al.\upcite{r24} proposed a novel IRS-UAV wireless communication scheme to maximize the spectrum efficiency and the energy efficiency via jointly optimizing the active beamforming, the passive beamforming and UAV trajectory. To the best of our knowledge, studying the cooperative trajectory optimization, computation offloading and resource allocation in the dynamic environment remains an open question, and moreover, determining the appropriate balance between the flying subslot and hovering subslot for efficient computation services is still an unresolved challenge.

\section{System Model}
\subsection{Network Model}
\begin{figure}[t]
  \begin{center}
  \includegraphics[width=3.3in]{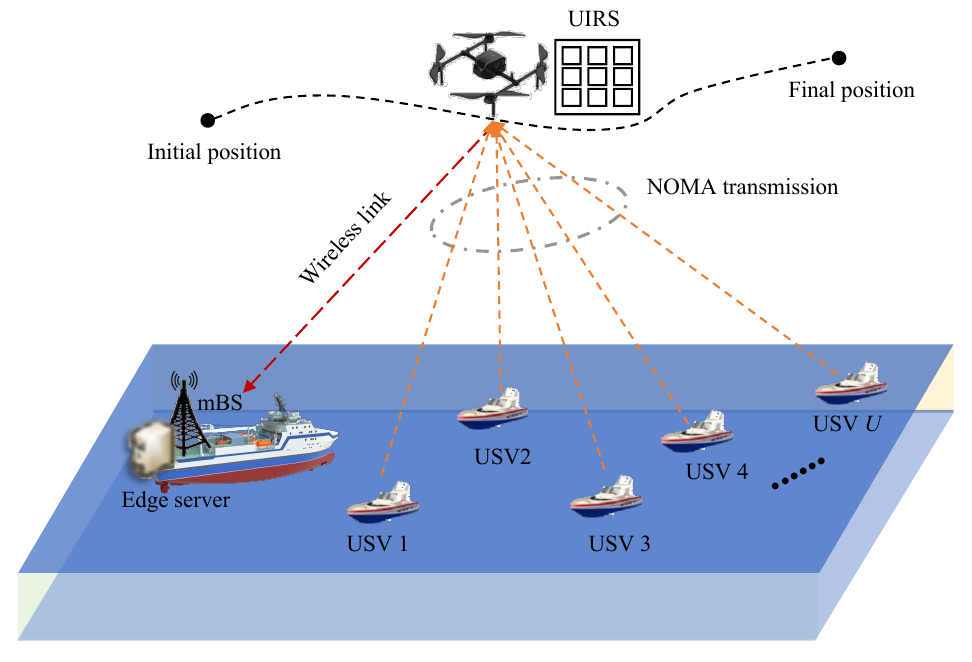}\\
  \caption{\textbf{The UIRS-assisted maritime edge computing network.}}\label{circuit_diagram}
  \end{center}
\end{figure}
In this paper, we consider a UIRS-assisted maritime edge computing network that consists of a group of USVs with a single antenna denoted as $\mathcal{U}=\{1,2,...,U\}$, a single antenna Micro-BS (mBS) and a UIRS, as shown in Fig. 1. Each USV has a computation processor to compute its tasks locally. Since the computation capability and battery capacity of USV are limited, the USV can be associated with the mBS for computation offloading. The mBS co-deployed with the edge server provides communication and computation services for USVs within a considered sea area. To facilitate the trajectory design, we divide the time span $\tau$ into $K$ time slots with identical lengths denoted as $\Delta = {\tau}/{K}$. Each time slot $k \in \mathcal{K}=\{1,2,...,K\}$ is partitioned into the flying time $\Delta_k^{\rm fly}$ and the hovering time $\Delta_k^{\rm hov}$, satisfying

\begin{equation}
    0 \leq \Delta_k^{\rm fly} \leq \Delta, 0 \leq \Delta_k^{\rm hov} \leq \Delta, \forall k \in \mathcal{K},
\end{equation}
\begin{equation}
    \Delta_k^{\rm fly}+\Delta_k^{\rm hov}=\Delta.
\end{equation}

\begin{table}
\renewcommand{\arraystretch}{1.0}
	\caption{\textbf{Main Notations and Descriptions}}
	\begin{center}
		\begin{tabular}{c p{6.3cm}}
		\hline
		\textbf{Notation} & \textbf{Description} \\
        \hline
		$U$ & The number of USVs\\
        $K$ & The number of time slots \\
        $N$ & The number of UIRS reflection elements\\
        $\Delta_k^{\rm fly}$ & Flying subslot in time slot $k$\\
        $\Delta_k^{\rm hov}$ & Hovering subslot in time slot $k$\\
        $\mathbf{l}_{u}$ & Coordinates of the USV $u$\\
        $\mathbf{l}_{\rm mBS}$ & Coordinates of the mBS\\
        $\mathbf{l}_{{\rm uirs},k}$ & Coordinates of UIRS in time slot $k$\\
        $v_k^{\rm hor}$& Horizontal flying speed of UIRS in time slot $k$\\
        $v_k^{\rm ver}$& Vertical flying speed of UIRS in time slot $k$\\
        $\varphi_{n,k}$ & Phase shift of $n$th reflection element in time slot $k$\\
        $\mathbf{h}_{u,k}^{\rm UI}$ & Channel gain between USV $u$ and the UIRS in time slot $k$\\
        $\mathbf{h}_{k}^{\rm IB}$ & Channel gain between UIRS and mBS in time slot $k$\\
        $h_{u}^{\rm UB}$ & Channel gain between USV $u$ and mBS\\
        $d_{u,k}^{\rm UI}$ & Distance between USV $u$ and UIRS in time slot $k$\\
        $d_k^{\rm IB}$ & Distance between UIRS and mBS in time slot $k$\\
        $d_{u}^{\rm UB}$ & Distance between USV $u$ and mBS\\
        $H_u$ & Antenna height of USV $u$\\
        $H_{\rm mBS}$ & Antenna height of mBS\\
        $R_{u,k}$ & The data rate of USV $u$ in time slot $k$\\
        $W$ & Channel bandwidth\\
        $p_{u,k}$ & Transmit power of USV $u$ in time slot $k$\\
        $p_{\rm hov}$ & Hovering power of UIRS\\
        $\alpha_{u,k}$ & Offloading ratio of USV $u$ in time slot $k$ \\
        $C_u^{\rm loc}$ & Required CPU cycles per bit at USV $u$ \\
        $C_{\rm es}$ & Required CPU cycles per bit at edge server\\
        $b_{u,k}$ & Number of bits generated by USV $u$ in time slot $k$\\
        $F_u$ & Computation capability of USV $u$\\
        $F_{u,k}^{\rm es}$ & Computation capability of edge server allocated to serve USV $u$ in time slot $k$\\
        \hline
		\end{tabular}
		\label{tab1}
	\end{center}
\end{table}

The UIRS with $N=N_x \times N_y$ reflection elements is deployed to assist the computation offloading. The reflection-coefficient matrix of the UIRS in the $k$th time slot is denoted by $\boldsymbol{\Theta}_k=diag\{e^{j\varphi_{1,k}},e^{j\varphi_{2,k}},...,e^{j\varphi_{N,k}}\}, \forall n \in \mathcal{N} \triangleq \{1,2,...,N\}$, where $\varphi_{n,k} \in [0, 2\pi]$ denotes the programmable phase shift of the $n$th reflection element and $j$ denotes the imaginary unit. We adopt NOMA to serve multiple USVs on the same frequency channel simultaneously through the non-orthogonal bandwidth resource for improving the spectrum efficiency. The main notations are summarized in Table I.

In the 3-Dimensional (3D) cartesian coordinate system, the coordinate of the USV $u$ is denoted as $\mathbf{l}_{u} = [x_{u},y_{u},0]^{\rm T}, \forall u \in \mathcal{U}$, where the Z-axis coordinate of the sea surface is zero. The coordinate of the mBS is denoted as $\mathbf{l}_{\rm mBS}=[x_{\rm mBS}, y_{\rm mBS}, z_{\rm mBS}]^{\rm T}$, where $z_{\rm mBS}$ denotes the height above the sea surface. The coordinate of the UIRS in the $k$th time slot is denoted as $\mathbf{l}_{{\rm uirs},k}=[x_{{\rm uirs},k},y_{{\rm uirs},k}, z_{{\rm uirs},k}]^{\rm T}, \forall k \in \mathcal{K}$, where $z_{{\rm uirs},k}$ denotes the altitude height above the sea surface in the $k$th time slot. In the $k$th time slot, the distance between the USV $u$ and the UIRS denotes $d_{u,k}^{\rm UI}={\|\mathbf{l}_{{\rm uirs},k}-\mathbf{l}_u\|}$, the distance between the UIRS and the mBS denotes $d_{k}^{\rm IB}={\|\mathbf{l}_{{\rm uirs},k}-\mathbf{l}_{\rm mBS}\|}$ and the distance between the USV $u$ and the mBS denotes $d_{u}^{\rm UB}={\|\mathbf{l}_{\rm mBS}-\mathbf{l}_u\|}$.

We assume that the location of UIRS remains unchanged within each time slot. The horizontal flying speed and vertical flying speed of the UIRS in the $k$th time slot satisfy following flying constraints
\begin{subequations}
	\begin{align}
		v_k^{\rm hor}= &\frac{\sqrt{(x_{{\rm uirs},k+1}-x_{{\rm uirs},k})^2+(y_{{\rm uirs},k+1}-y_{{\rm uirs},k})^2}}{\Delta_k^{\rm fly}}, \notag \\&
        0 \leq v_k^{\rm hor} \leq v_{\rm max}^{\rm hor}, \forall k \in \mathcal{K}, \\
		v_k^{\rm ver}= &\frac{|z_{{\rm uirs},k+1}-z_{{\rm uirs},k}|}{\Delta_k^{\rm fly}}, 0 \leq v_k^{\rm ver} \leq v_{\rm max}^{\rm ver}, \forall k \in \mathcal{K},
	\end{align}
\end{subequations}
where $v_{\rm max}^{\rm hor}$ denotes the maximum horizontal flying speed of UIRS and $v_k^{\rm ver}$ denotes the maximum vertical flying speed of UIRS.
\subsection{Communication Model}
We assume that links from USVs to the UIRS (U-I link) and the UIRS to the mBS (I-B link) follow LoS. In the $k$th time slot, the channel gain $\mathbf{h}_{u,k}^{\rm UI} \in \mathbb{C}^{N \times 1}$ between the USV $u$ and the UIRS can be given as\upcite{r40}
\begin{equation}
\begin{aligned}
	& \mathbf{h}_{u,k}^{\rm UI}= \\
	& \sqrt{\rho_0 (d_{u,k}^{\rm UI})^{-\beta}}[1,e^{-j\frac{2\pi d}{\lambda}{\rm sin}\phi_{u,k}^{\rm UI}{\rm cos}\gamma_{u,k}^{\rm UI}},...,\\
    &\quad e^{-j\frac{2(N_x-1)\pi d}{\lambda}{\rm sin}\phi_{u,k}^{\rm UI}{\rm cos}\gamma_{u,k}^{\rm UI}}]^{\rm T} \otimes [1,e^{-j\frac{2\pi d}{\lambda}{\rm sin}\phi_{u,k}^{\rm UI}{\rm sin}\gamma_{u,k}^{\rm UI}}\\
    &\quad ,...,e^{-j\frac{2(N_y-1)\pi d}{\lambda}{\rm sin}\phi_{u,k}^{\rm UI}{\rm sin}\gamma_{u,k}^{\rm UI}}]^{\rm T}, \forall u \in \mathcal{U}, \forall k \in \mathcal{K},
\end{aligned}
\end{equation}
where $\rho_0$ denotes the channel gain at the reference distance $d_0=1 \rm m$, $\beta \geq 2$ denotes the path loss exponent of the U-I link, $\lambda$ denotes the carrier frequency wavelength, $d$ denotes the separation distance. The vertical/horizontal Angle of Arrivals (AoAs) from the USV to the UIRS and the vertical/horizontal of Angle of Departure (AoDs) from the UIRS to the mBS are shown in Fig.2. $\phi_{u,k}^{\rm UI}$ and $\gamma_{u,k}^{\rm UI}$ denote the vertical and horizontal AoAs from the USV $u$ to the UIRS in the $k$th time slot, respectively. We have ${\rm sin}\phi_{u,k}^{\rm UI}=\frac{|y_{{\rm uirs},k}-y_{u}|}{d_{u,k}^{\rm UI}}$, ${\rm sin}\gamma_{u,k}^{\rm UI}=\frac{|x_{{\rm uirs},k}-x_{u}|}{\sqrt{(x_{{\rm uirs},k}-x_{u})^2+(y_{{\rm uirs},k}-y_{u})^2}}$ and ${\rm cos}\gamma_{u,k}^{\rm UI}=\frac{|y_{{\rm uirs},k}-y_{u}|}{\sqrt{(x_{{\rm uirs},k}-x_{u})^2+(y_{{\rm uirs},k}-y_{u})^2}}$.
\begin{figure}[t]
\centering
    \subfigure[The vertical AoAs and AoDs]{
       \centering
        \includegraphics[width=0.22\textwidth]{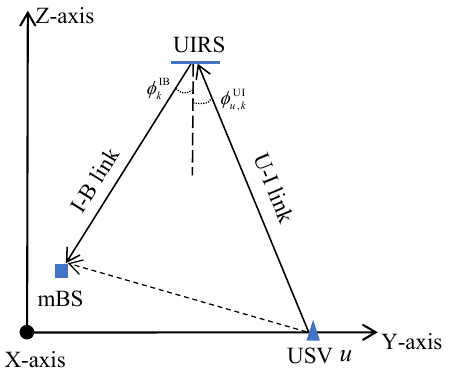}
    }\subfigure[The horizontal AoAs and AoDs]{
        \centering
        \includegraphics[width=0.22\textwidth]{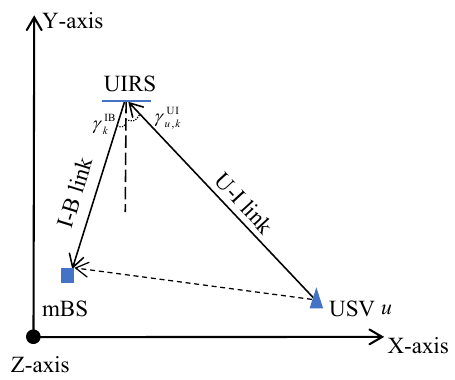}
    }\caption{\textbf{The vertical/horizontal AoAs and AoDs.}\label{inut_output}}
\end{figure}

In the $k$th time slot, the channel gain $\mathbf{h}_k^{\rm IB} \in \mathbb{C}^{N \times 1}$ between the UIRS and the mBS can be expressed as\upcite{r40}
\begin{equation}
\begin{aligned}
	\mathbf{h}_{k}^{\rm IB}
	& = \sqrt{\rho_0 (d_{k}^{\rm IB})^{-\beta}}[1,e^{-j\frac{2\pi d}{\lambda}{\rm sin}\phi_{k}^{\rm IB}{\rm cos}\gamma_{k}^{\rm IB}},...,\\
    &\quad e^{-j\frac{2(N_x-1)\pi d}{\lambda}{\rm sin}\phi_{k}^{\rm IB}{\rm cos}\gamma_{k}^{\rm IB}}]^{\rm T} \otimes [1,e^{-j\frac{2\pi d}{\lambda}{\rm sin}\phi_{k}^{\rm IB}{\rm sin}\gamma_{k}^{\rm IB}}\\
    &\quad ,...,e^{-j\frac{2(N_y-1)\pi d}{\lambda}{\rm sin}\phi_{k}^{\rm IB}{\rm sin}\gamma_{k}^{\rm IB}}]^{\rm T}, \forall k \in \mathcal{K},
\end{aligned}
\end{equation}
where $\phi_{k}^{\rm IB}$ and $\gamma_{k}^{\rm IB}$ denote the vertical and horizontal AoDs from the UIRS to the mBS in the $k$th time slot, respectively. We have ${\rm sin}\phi_{k}^{\rm IB}=\frac{|y_{{\rm uirs},k}-y_{\rm mBS}|}{d_{k}^{\rm IB}}$, ${\rm sin}\gamma_{k}^{\rm IB}=\frac{|x_{{\rm uirs},k}-x_{\rm mBS}|}{\sqrt{(x_{{\rm uirs},k}-x_{\rm mBS})^2+(y_{{\rm uirs},k}-y_{\rm mBS})^2}}$ and ${\rm cos}\gamma_{k}^{\rm IB}=\frac{|y_{{\rm uirs},k}-y_{\rm mBS}|}{\sqrt{(x_{{\rm uirs},k}-x_{\rm mBS})^2+(y_{{\rm uirs},k}-y_{\rm mBS})^2}}$.

Due to impacts of sea surface reflection and antenna height, we consider the two-ray reflection model for large-scale fading of maritime channels\upcite{r3}. The channel gain of the direct link $h_{u}^{\rm UB} \in \mathbb{C}^{1 \times 1}$ between the USV $u$ and the mBS can be expressed as\upcite{r59}
\begin{equation}
		h_{u}^{\rm UB}
		 =\left(\frac{\lambda}{4 \pi d_{u}^{\rm UB}}\right)^2\left[2\sin\left(\frac{2\pi H_u H_{\rm mBS}}{\lambda d_{u}^{\rm UB}}\right)\right]^2,
		 \forall u \in \mathcal{U},
\end{equation}
where $H_u$ denotes the antenna height of the USV $u$, $H_{\rm mBS}$ denotes the antenna height of the mBS.

The NOMA protocol is used for computation offloading from USVs to the UIRS, which adopts the Successive Interference Cancellation (SIC) to mitigate the co-channel interference. In this way, the channel gains of USVs for the cascaded USV-UIRS-mBS link sorted in the ascending ordering in the $k$th time slot can be expressed as\upcite{r53,r31}
\begin{equation}
	\begin{aligned}
	&\left|h_{1}^{\rm UB}+(\mathbf{h}_k^{\rm IB})^{\rm H}\boldsymbol{\Theta}_k \mathbf{h}_{1,k}^{\rm UI}\right|^2 \leq \cdots  \leq \\
    &\quad \left|h_{U}^{\rm UB}+(\mathbf{h}_k^{\rm IB})^{\rm H}\boldsymbol{\Theta}_k \mathbf{h}_{U,k}^{\rm UI}\right|^2.
	\end{aligned}
\end{equation}

The signal received from the USV with the highest channel gain would be decoded first, while other USVs with lower channel gains would be regarded as co-channel interference. Thus, the data rate of USV $u$ in the $k$th time slot can be expressed as 
\begin{equation}
	\begin{aligned}
	& R_{u,k}= \\
	& W\log_{2}\left(1+\frac{p_{u,k}\left|h_{u}^{\rm UB}+(\mathbf{h}_k^{\rm IB})^{\rm H}\boldsymbol{\Theta}_k \mathbf{h}_{u,k}^{\rm UI}\right|^2}{\sum\limits_{i=1}^{u-1}{p_{i,k}\left|h_{i}^{\rm UB}+(\mathbf{h}_k^{\rm IB})^{\rm H}\boldsymbol{\Theta}_{k} \mathbf{h}_{i,k}^{\rm UI}\right|^2}+\sigma^2}\right)\\
	& \quad ,\forall u \in \mathcal{U}, \forall k \in \mathcal{K},
	\end{aligned}
\end{equation}
where $W$ denotes the channel bandwidth, $\sigma^2$ denotes the additive Gaussian noise power, $p_{u,k}$ denotes the transmit power of USV $u$. Considering the practical limit of the radio interface\upcite{r26}, USV $u$ has a maximum power $P_u^{\rm max}$ constraint, satisfying
\begin{equation}
	0 \leq p_{u,k} \leq P_u^{\rm max}, \forall u \in \mathcal{U}, \forall k \in \mathcal{K}.
\end{equation}
\subsection{Energy Consumption Model}
In the $k$th time slot, $b_{u,k}$-bit computation tasks generated by USV $u$ should be processed locally and offloaded to the mBS's edge server based on the partial offloading strategy. Let $\alpha_{u,k} \in [0,1]$ denote the ratio of data size offloaded from the USV $u$ to the edge server, and $(1-\alpha_{u,k}) \in [0,1]$ denotes the ratio of data size processed by USV $u$ locally in the $k$th time slot. It is noted that the edge server and USVs process computation tasks individually. We consider that the UIRS is required to keep a hovering state to perform signal reflection, and thus the hovering time is equal to the computation offloading time. Accordingly, the offloaded data size of each USV to be processed cannot exceed its maximum transmission data size, satisfying
\begin{equation}
	\alpha_{u,k}b_{u,k} \leq R_{u,k}\Delta_k^{\rm hov}, \forall u \in \mathcal{U}, \forall k \in \mathcal{K}.
\end{equation}

The energy consumed by the UIRS reflection is negligible since the IRS passive beamforming consumes ultra-low power in tuning the UIRS phase shifts. The local computation energy consumption of USV $u$ in the $k$th time slot can be expressed as
\begin{equation}
	E_{u,k}^{\rm loc}=\kappa_u(1-\alpha_{u,k})b_{u,k}C_u^{\rm loc}(F_u)^2, \forall u \in \mathcal{U}, \forall k \in \mathcal{K},
\end{equation}
where $C_u^{\rm loc}$ denotes the number of cycles required to finish one bit in the USV $u$, $F_u$ denotes the computation capability of USV $u$, $\kappa_u$ denotes the switched capacitance coefficient for USV $u$. The transmission energy consumption of USV $u$ in the $k$th time slot can be expressed as
\begin{equation}
	E_{u,k}^{\rm trans}=\Delta_k^{\rm hov}p_{u,k}, \forall u \in \mathcal{U}, \forall k \in \mathcal{K}.
\end{equation}

The computation energy consumption of the edge server for computing data from the USV $n$ in the $k$th time slot is expressed as
\begin{equation}
	E_{u,k}^{\rm es}=\kappa_{\rm es}\alpha_{u,k}b_{u,k}C_{\rm es}(F_{u,k}^{\rm es})^2, \forall u \in \mathcal{U}, \forall k \in \mathcal{K},
\end{equation}
where $C_{\rm es}$ denotes the number of cycles required to finish one bit in the edge server, $F_{u,k}^{\rm es}$ denotes the edge computation capability allocated to serve the USV $u$ in the $k$th time slot, $\kappa_{\rm es}$ denotes the switched capacitance coefficient for the edge server. The edge computation capability allocated to serve the USV $u$ is subject to the maximum computation capability $F_{\rm es}^{\rm max}$, satisfying
\begin{equation}
    	\sum_{u=1}^{U}F_{u,k}^{\rm es} \leq F_{\rm es}^{\rm max}, \forall k \in \mathcal{K}.
\end{equation}

We adopt the energy consumption model of rotary-wing UAV, considering the horizontal and vertical velocities. According to \cite{r61,r62}, the flying energy consumption in the $k$th time slot is expressed as
\begin{equation}
	\begin{aligned}
		  & E_k^{\rm fly} = 
        \Delta_k^{\rm fly}\Bigg(\frac{1}{2}\xi \rho S_0A_0(v_k^{\rm hor})^3 +  P_0\bigg(1+\frac{3(v_k^{\rm hor})^2}{U_{\rm tip}^2}\bigg) \Bigg.\\ 
		&\Bigg. + P_1 \Bigg(\sqrt{(1+\frac{(v_k^{\rm hor})^4}{4(v^{\prime})^4}} -\frac{(v_k^{\rm hor})^2}{2(v^{\prime})^2}\Bigg)^{\frac{1}{2}}+ P_2v_k^{\rm ver} \Bigg)\\
        &, \forall k \in \mathcal{K},
	\end{aligned}
\end{equation}
where $P_0$ denotes the blade profile power, $P_1$ denotes the induced power, $P_2$ denotes the power throughout descent and ascent, $U_{\rm tip}$ denotes the tip velocity of the rotor blade, $\xi$ denotes the fuselage drag ratio, $S_0$ denotes rotor solidity, $\rho$ denotes the air density, $A_0$ denotes the rotor disc area, $v^{\prime}$ denotes rotor induced velocity in hovering.

To perform signal reflection, the UIRS is required to maintain a hovering state, and thus the hovering energy consumption in the $k$th time slot can be expressed as
\begin{equation}
	E_k^{\rm hov} = p_{\rm hov}\Delta_k^{\rm hov}, \forall k \in \mathcal{K},
\end{equation}
where $p_{\rm hov}$ denotes the hovering power of UIRS.

Based on the above models, the weighted overall energy consumption of the transmission and computation energy consumption of USVs, the computation energy consumption of the edge server, the flying and hovering energy consumption of the UIRS can be expressed as
\begin{equation}
	\begin{aligned}
	E^{\rm tot}
	& =\sum_{k=1}^{K}\sum_{u=1}^{U}(E_{u,k}^{\rm loc}+E_{u,k}^{\rm trans})+\sum_{k=1}^{K}\sum_{u=1}^{U}E_{u,k}^{\rm es} + \\
	&\quad \eta\sum_{k=1}^{K}(E_k^{\rm fly}+E_k^{\rm hov}),
\end{aligned}
\end{equation}
where $\eta$ denotes the weight factor.
\subsection{Problem Formulation}
We propose an overall energy consumption minimization problem by jointly optimizing subslot duration, offloading ratios of USVs, transmit power of USVs, computation capability allocation, UIRS phase shifts and 3D trajectory of the UIRS, with the objective of minimizing the overall energy consumption in processing tasks generated by USVs for UIRS-assisted maritime edge computing networks. Let $\boldsymbol{\Delta}^{\rm fly} = \{\Delta_k^{\rm fly}, \forall k \in \mathcal{K}\}$ denote the flying subslot, $\boldsymbol{\Delta}^{\rm hov} = \{\Delta_k^{\rm hov}, \forall k \in \mathcal{K}\}$ denote the hovering subslot, $\boldsymbol{\alpha}=\{\alpha_{u,k},\forall u \in \mathcal{U}, \forall k \in \mathcal{K}\}$ denote the offloading ratios of USVs, $\mathbf{F}=\{F_{u,k}^{\rm es}, \forall u \in \mathcal{U}, \forall k \in \mathcal{K}\}$ denote the computation capability allocation, $\mathbf{p}=\{p_{u,k},\forall u \in \mathcal{U}, \forall k \in \mathcal{K}\}$ denote the transmit power of USVs, $\boldsymbol{\varphi}=\{\varphi_{n,k}, \forall n \in \mathcal{N}, \forall k \in \mathcal{K}\}$ denote the UIRS phase shifts, and $\mathbf{L}=\{\mathbf{l}_{{\rm uirs},k},\forall k \in \mathcal{K}\}$ denote the UIRS trajectory. The overall energy consumption minimization problem \textbf{P1} can be formulated as
\begin{subequations}
\begin{align*}
    {\rm (\mathbf{P1}):}
	&\min_{\boldsymbol{\Delta}^{\rm fly},\boldsymbol{\Delta}^{\rm hov}\boldsymbol{\alpha},\mathbf{F},\mathbf{p},\boldsymbol{\varphi},\mathbf{L}}\
	E^{\rm tot} \tag{18}\\
	\text{s.t.} &\quad 0 \leq \varphi_{n,k} \leq 2\pi,\forall n \in \mathcal{N},\forall k \in \mathcal{K}, \tag{18a}\\
	&\quad 0 \leq \alpha_{u,k} \leq 1, \forall u \in \mathcal{U}, \forall k \in \mathcal{K}, \tag{18b}\\
	&\quad	\sum_{u=1}^{U}F_{u,k}^{\rm es} \leq F_{\rm es}^{\rm max}, \forall k \in \mathcal{K}, \tag{18c}\\
	&\quad 0 \leq F_{u,k}^{\rm es} \leq F_{\rm es}^{\rm max}, \forall u \in \mathcal{U}, \forall k \in \mathcal{K}, \tag{18d}\\
	&\quad 0 \leq p_{u,k} \leq P_u^{\rm max}, \forall u \in \mathcal{U}, \forall k \in \mathcal{K}, \tag{18e}\\
	&\quad \alpha_{u,k}b_{u,k} \leq R_{u,k}\Delta_k^{\rm hov}, \forall u \in \mathcal{U}, \forall k \in \mathcal{K}, \tag{18f}\\
    &\quad 0 \leq \Delta_k^{\rm fly} \leq \Delta, 0 \leq \Delta_k^{\rm hov} \leq \Delta, \forall k \in \mathcal{K}, \tag{18g}\\
    &\quad \Delta_k^{\rm fly}+\Delta_k^{\rm hov}=\Delta, \forall k \in \mathcal{K}, \tag{18h}\\
	&\quad 0 \leq v_k^{\rm hor} \leq v_{\rm max}^{\rm hor}, \forall k \in \mathcal{K}, \tag{18i}\\
	&\quad 0 \leq v_k^{\rm ver} \leq v_{\rm max}^{\rm ver}, \forall k \in \mathcal{K}. \tag{18j}
\end{align*}
\end{subequations}
Constraint (18a) ensures the range of the UIRS phase shift. Constraint (18b) ensures the scope of USV's offloading ratio. Constraints (18c) and (18d) limit the range of the allocated edge computation capability. Constraint (18e) ensures that the transmit power of USV cannot exceed the maximum power. Constraint (18f) ensures that the offloaded data size of each USV to be processed cannot exceed its maximum transmission data size. Constraints (18g) and (18h) are the boundary constraints of the duration of flying subslot and hovering subslot. Constraints (18i) and (18j) guarantee that the horizontal flying speed and vertical flying speed of UIRS should be less than the maximum speed $v_{\rm max}^{\rm hor}$ and $v_{\rm max}^{\rm ver}$, respectively.

\section{Algorithm Design}
The \textbf{P1} is a non-convex problem that is difficult to solve directly and effectively due to the coupling of optimization variables. Therefore, we decompose the problem \textbf{P1} into two tractable subproblems. The details of the proposed RMTO scheme are summarized in \textbf{Algorithm 1}. Specifically, in the RMTO-Top optimization stage, we derive a closed-form optimal solution for the transmit power of USVs and employ the SDR method to optimize the UIRS phase shifts via \textbf{Algorithm 2}. Then, in the RMTO-Sub optimization stage, we optimize the subslot duration, offloading ratios of USVs, computation capability allocation and UIRS trajectory via \textbf{Algorithm 3}. Through a certain number of iterations, the optimal solution can be obtained eventually.
\begin{algorithm}[t]
        \caption{\textbf{RMTO Scheme}}
        \LinesNumbered
        \KwIn{Maximum steps $K$.}
        \KwOut{An optimal solution $\{\boldsymbol{\Delta}^{\rm fly}, \boldsymbol{\Delta}^{\rm hov}, \boldsymbol{\alpha}, \mathbf{F}, \mathbf{p}, \boldsymbol{\varphi}, \mathbf{L}\}$ to \textbf{P1}.}
		{Initialize $\{\boldsymbol{\Delta}^{\rm fly}_0, \boldsymbol{\Delta}^{\rm hov}_0, \boldsymbol{\alpha}_0, \mathbf{F}_0, \mathbf{L}_0\}$.}
  
        \For{step = $k \in \{1,2,...,K\}$}{
        
        {Given $\{\boldsymbol{\Delta}^{\rm fly}_k, \boldsymbol{\Delta}^{\rm hov}_k, \boldsymbol{\alpha}_k, \mathbf{F}_k, \mathbf{L}_k\}$, optimize the transmit power and the UIRS phase shifts $\{\boldsymbol{\varphi}_{k+1}, \mathbf{p}_{k+1}\}$ by RMTO-Top (\textbf{Algorithm 2}).}
        
        {Given $\{\boldsymbol{\varphi}_{k+1}, \mathbf{p}_{k+1}\}$, optimize the subslot duration, offloading ratios, computation capability allocation and UIRS trajectory $\{\boldsymbol{\Delta}^{\rm fly}_{k+1}, \boldsymbol{\Delta}^{\rm hov}_{k+1}, \boldsymbol{\alpha}_{k+1}, \mathbf{F}_{k+1}, \mathbf{L}_{k+1}\}$ by RMTO-Sub (\textbf{Algorithm 3}).}
        }
        
        \textbf{Return} $\boldsymbol{\Delta}^{\rm fly}, \boldsymbol{\Delta}^{\rm hov}, \boldsymbol{\alpha}, \mathbf{F}, \mathbf{p}, \boldsymbol{\varphi}, \mathbf{L}$.
\end{algorithm} 
\subsection{RMTO-Top Optimization}
In this subsection, supposing that the subslot duration $\boldsymbol{\Delta}^{\rm fly}$, $\boldsymbol{\Delta}^{\rm hov}$, the offloading ratios of USVs $\boldsymbol{\alpha}$, computation capability allocation $\mathbf{F}$, and UIRS trajectory $\mathbf{L}$ are given in advance. We first derive the transmit power of USVs in closed-form, and then focus on the UIRS phase shifts optimization. According to (18f), we can get $\frac{\alpha_{u,k}b_{u,k}}{\Delta_k^{\rm hov}} \leq R_{u,k}$. Combining with (8), we can obtain the following equality as
\begin{equation}
	\begin{aligned}
		p_{u,k}
		& =\frac{\sum_{i=1}^{u-1}p_{i,k}\left|h_{i}^{\rm UB}+(\mathbf{h}_k^{\rm IB})^{\rm H}\boldsymbol{\Theta}_k \mathbf{h}_{i,k}^{\rm UI})\right|^2+\sigma^2}{\left|h_{u}^{\rm UB}+(\mathbf{h}_k^{\rm IB})^{\rm H}\boldsymbol{\Theta}_k \mathbf{h}_{u,k}^{\rm UI})\right|^2} \\ & \quad \times \left(e^{\frac{\ln2 \alpha_{u,k}b_{u,k}}{W \Delta_k^{\rm hov}}}-1\right),
		\forall u \in \mathcal{U}, \forall k \in \mathcal{K}.
	\end{aligned}
\end{equation}

For simplicity, we define
\begin{equation}
	\left\{
	\begin{aligned}
		&J=\frac{\ln2}{W}, \\
		&S_{u,k}=\sum_{i=1}^{u}p_{i,k}\left|h_{i}^{\rm UB}+(\mathbf{h}_k^{\rm IB})^{\rm H}\boldsymbol{\Theta}_k \mathbf{h}_{i,k}^{\rm UI})\right|^2+\sigma^2.\\
	\end{aligned}
	\right.
	\label{eq:1}
\end{equation}

According to the recursive form, (20) can be rewritten as
\begin{equation}
	\begin{aligned}
		S_{u,k}=e^{J\frac{\alpha_{u,k}b_{u,k}}{\Delta_k^{\rm hov}}}S_{u-1,k}, \forall u \in \mathcal{U}\setminus \{1\}.
	\end{aligned}
\end{equation}

Combining with (20) and (21), we can get that $S_{1,k}=p_{1,k}\left|h_{1}^{\rm UB}+(\mathbf{h}_k^{\rm IB})^{\rm H}\boldsymbol{\Theta}_k \mathbf{h}_{1,k}^{\rm UI})\right|^2+\sigma^2=\sigma^2 e^{J\frac{\alpha_{1,t}b_{1,k}}{\Delta_k^{\rm hov}}}$. Then, we can obtain the general expression as
\begin{equation}
	\begin{aligned}
		S_{u,k}=
		& S_{1,k}e^{J\sum_{i=2}^{u}\frac{\alpha_{i,k}b_{i,k}}{\Delta_k^{\rm hov}}}=\sigma^2e^{J\sum_{i=1}^{u}\frac{\alpha_{i,k}b_{i,k}}{\Delta_k^{\rm hov}}}, \\
		& \forall u \in \mathcal{U}, \forall k \in \mathcal{K}.
	\end{aligned}
\end{equation}

Thus, the optimal transmit power of the USV $u$ in the $k$th time slot can be transformed as
\begin{equation}
	\begin{aligned}
		& p_{u,k}=
		\frac{S_{u,k}-S_{u-1,k}}{\left|h_{u}^{\rm UB}+(\mathbf{h}_k^{\rm IB})^{\rm H}\boldsymbol{\Theta}_k \mathbf{h}_{u,k}^{\rm UI})\right|^2} \\
        &\quad\quad = \frac{\sigma^2(e^{Jx_{u,k}}-e^{Jx_{u-1,k}})}{\left|h_{u}^{\rm UB}+(\mathbf{h}_k^{\rm IB})^{\rm H}\boldsymbol{\Theta}_k \mathbf{h}_{u,k}^{\rm UI})\right|^2}, \forall u \in \mathcal{U}, \forall k \in \mathcal{K},
	\end{aligned}
\end{equation}
where $x_{u,k}=\sum_{i=1}^{u}\frac{\alpha_{i,k}b_{i,k}}{\Delta_k^{\rm hov}}$. Problem \textbf{P1} can be rewritten as
\begin{subequations}
	\begin{align*}
		{\rm (\mathbf{P2}):}
	    &\min_{\boldsymbol{\varphi}}\ \sum_{k=1}^{K}\sum_{u=0}^{U-1}(\frac{\sigma^2}{\left|h_{u}^{\rm UB}+(\mathbf{h}_k^{\rm IB})^{\rm H}\boldsymbol{\Theta}_k \mathbf{h}_{u,k}^{\rm UI})\right|^2}\\
		&-\frac{\sigma^2}{\left|h_{u+1}^{\rm UB}+(\mathbf{h}_k^{\rm IB})^{\rm H}\boldsymbol{\Theta}_k \mathbf{h}_{U+1,k}^{\rm UI})\right|^2})\times e^{Jx_{u,k}}\Delta_k^{\rm hov} \tag{24}\\
		\text{s.t.}& \quad e^{Jx_{u,k}}-e^{Jx_{u-1,k}} \leq \\
        &\quad \frac{P_u^{\rm max}\left|h_{u}^{\rm UB}+(\mathbf{h}_k^{\rm IB})^{\rm H}\boldsymbol{\Theta}_k \mathbf{h}_{u,k}^{\rm UI})\right|^2}{\sigma^2},\\
        &\quad \forall u \in \mathcal{U}, \forall k \in \mathcal{K}, \tag{24a}\\
		&\quad 0 \leq \varphi_{n,k} \leq 2\pi, \forall n \in \mathcal{N}, \forall k \in \mathcal{K}. \tag{24b}\\
	\end{align*}
\end{subequations}

Let $\mathbf{w}_k=[w_{1,k},...,w_{N,k}]$, where $w_{n,k}=e^{j\varphi_{n,k}}$. By employing the transformation of variables we have $\left|h_{u}^{\rm UB}+(\mathbf{h}_k^{\rm IB})^{\rm H}\boldsymbol{\Theta}_k \mathbf{h}_{u,k}^{\rm UI})\right|^2=\left|h_{u}^{\rm UB}+\boldsymbol{\Phi}_{u,k}\mathbf{w}_k^{\rm H}\right|^2$, where $\boldsymbol{\Phi}_{u,k}=(\mathbf{h}_k^{\rm IB})^{\rm H} diag(\mathbf{h}_{u,k}^{\rm UI})$. Constraint (24b) can be rewritten as $|w_{n,k}|=1, \forall n \in \mathcal{N}, \forall k \in \mathcal{K}$. Thus, we can transform the problem \textbf{P2} into the following equivalent form
\begin{subequations}
	\begin{align*}
		{\rm (\mathbf{P3}):}
	    &\min_{\mathbf{w}_k}\ \sum_{k=1}^{K}\sum_{u=0}^{U-1}(\frac{\sigma^2}{\left|h_{u}^{\rm UB}+\boldsymbol{\Phi}_{u,k}\mathbf{w}_k^{\rm H}\right|^2}\\
		&\quad\quad - \frac{\sigma^2}{\left|h_{u+1}^{\rm UB}+\boldsymbol{\Phi}_{u+1,k}\mathbf{w}_k^{\rm H}\right|^2}) \times e^{Jx_{u,k}}\Delta_k^{\rm hov} \tag{25}\\
		\text{s.t.}& \quad e^{Jx_{u,k}}-e^{Jx_{u-1,k}} \leq \frac{P_u^{\rm max}\left|h_{u}^{\rm UB}+\boldsymbol{\Phi}_{u,k}\mathbf{w}_k^{\rm H}\right|^2}{\sigma^2}\\
		&\quad ,\forall u \in \mathcal{U}, \forall k \in \mathcal{K}, \tag{25a}\\
		&\quad |w_{n,k}|=1, \forall n \in \mathcal{N}, \forall k \in \mathcal{K}. \tag{25b}\\
	\end{align*}
\end{subequations}

We further transform the problem \textbf{P3} into a convex optimization problem through the SDR method. By introducing the auxiliary variable $\varpi$, we have $\bar{\mathbf{w}}_k=\left[\begin{array}{c}\mathbf{w}_k^{\rm H} \\ \varpi \end{array}\right]$ and $rank(\mathbf{W}_k)=\bar{\mathbf{w}}_k \bar{\mathbf{w}}_k^{\rm H}$, satisfying $rank(\mathbf{W}_k)=1$ and $\mathbf{R}_{u,k}=\left[\begin{array}{cc}\Phi_{u,k}^{\rm H}\Phi_{u,k}&\Phi_{u,k}^{\rm H} h_{u}^{\rm UB} \\ h_{u}^{\rm UB}\Phi_{u,k}&0 \end{array}\right]$. Therefore, the problem \textbf{P3} can be reformulated as
\begin{subequations}
	\begin{align*}
		{\rm (\mathbf{P4}):}
	    &\min_{\mathbf{W}_k}\ \Delta_k^{\rm hov}\sum_{k=1}^{K}\sum_{u=0}^{U-1}\sigma^2(e^{Jx_{u,k}}-1)\frac{1}{tr(\mathbf{R}_{u,k}\mathbf{W}_k)}\tag{26}\\
		\text{s.t.}& \quad e^{Jx_{u,k}}-e^{Jx_{u-1,k}} \leq P_u^{\rm max}\frac{tr(\mathbf{R}_{u,k}\mathbf{W}_k)}{\sigma^2},\\
        &\quad \forall u \in \mathcal{U}, \forall k \in \mathcal{K}, \tag{26a}\\
		&\quad \mathbf{W}_{n,k}=1, \forall n \in \mathcal{N}, \forall k \in \mathcal{K}, \tag{26b}\\
		&\quad \mathbf{W}_k \succeq 0, \forall k \in \mathcal{K}. \tag{26c}\\
	\end{align*}
\end{subequations}

The problem \textbf{P4} is a convex optimization problem solved by CVXPY\upcite{r37}. Thus, the optimal value of $\mathbf{W}_k^*$ can be obtained. Then, we need the extra steps to convert $\mathbf{W}_k^*$ to a feasible solution of $\mathbf{w}_k^*$ according to the $\bar{\mathbf{w}}_k^*=\left[\begin{array}{c}({\mathbf{w}}_k^*)^{\rm H} \\ \varpi^* \end{array}\right]$ and $\mathbf{W}_k^*=\bar{\mathbf{w}}_k^*(\bar{\mathbf{w}}_k^*)^{\rm H}$. Finally, after optimized $\mathbf{w}_k^*$, we can get optimal UIRS phase shifts $\varphi_{n,k}^*$ according to $w_{n,k}^*=e^{j\varphi_{n,k}}$, and then get optimal transmit power of $u$th USV $p_{u,k}^*$ based on Eq. (23). The details of the RMTO-Top optimization is summarized in \textbf{Algorithm 2}.
\begin{algorithm}[t]
        \caption{\textbf{RMTO-Top Optimization}}
        \LinesNumbered
		\KwIn{Flying subslot $\boldsymbol{\Delta}^{\rm fly}_k$, hovering subslot $\boldsymbol{\Delta}^{\rm hov}_k$, offloading ratios of USVs $\boldsymbol{\alpha}_k$, computation capability allocation $\mathbf{F}_k$, UIRS trajectory $\mathbf{L}_k$ in the $k$th step.}
		\KwOut{UIRS phase shifts $\boldsymbol{\varphi}_{k}^*$, transmit power $\mathbf{p}_{k}^*$.}
        {Obtain $\mathbf{W}_k^*$ by solving Problem \textbf{P4}.}
        
        {Obtain $\mathbf{w}_k^*$ according to $\mathbf{W}_k^*$.}
        
        {Obtain $\varphi_{n,k}^*$ according to $w_{n,k}^*=e^{j\varphi_{n,k}}$.}

        {Compute $p_{u,k}^*$ based on the Eq. (23).}

		\textbf{Return} $\boldsymbol{\varphi}_k^*, \mathbf{p}_k^*$. 
\end{algorithm} 

\subsection{RMTO-Sub Optimization}
The DDPG is a policy-based and model-free deep reinforcement learning algorithm, which is considered to solve continuous variables in a dynamic environment. The underlying-layered subproblem is reformulated as an MDP. Specifically, the UIRS-assisted maritime edge computing scenario is treated as the environment entity and the mBS is treated as the agent. Besides, the key elements of MDP, such as state space $\mathcal{S}$, action space $\mathcal{A}$ and reward function $r$, are defined as follows.

$\bullet$ \textbf{State Space} $\mathcal{S}$: In the $k$th-round, the state space is expressed as $s_k=\{\mathbf{l}_{{\rm uirs},k},\mathbf{B}_k\}$. Here, $\mathbf{l}_{{\rm uirs},k}=[x_{{\rm uirs},k},y_{{\rm uirs},k},z_{{\rm uirs},k}]^{\rm T}$ means the 3D position of the UIRS, $\mathbf{B}_k=\{b_{1,k},b_{2,k},...,b_{U,k}\}$ means the size of task data generated by USVs.

$\bullet$ \textbf{Action Space} $\mathcal{A}$: In the $k$th-round, the action space is expressed as $a_k=\{\Delta_k^{\rm fly},\Delta_k^{\rm hov},\boldsymbol{\alpha}_k,\mathbf{F}_k,v_k^{\rm ver}, v_k^{\rm hor},\vartheta_k\}$. Here, $\Delta_k^{\rm fly}$ and $\Delta_k^{\rm hov}$ mean the flying subslot and hovering subslot, respectively, $\boldsymbol{\alpha}_k=\{\alpha_{1,k},\alpha_{2,k},...,\alpha_{U,k}\}$ means the offloading ratios of USVs, $\mathbf{F}_k=\{F_{1,k}^{es},F_{2,k}^{\rm es},...,F_{U,k}^{\rm es}\}$ means the computation capability allocation, $v_k^{\rm ver}\in [0,v_{\rm max}^{\rm ver}]$ means the vertical flying velocity of UIRS, $v_k^{\rm hor} \in [0,v_{\rm max}^{\rm hov}]$ means the horizontal flying velocity of UIRS, $\vartheta_k \in [0,2\pi)$ means the horizontal flying angle of UIRS.

$\bullet$ \textbf{Reward Function} $r$: The reward function is expressed as $r_k(s_k,a_k)=-\sum_{u=1}^{U}\left(E{_u^{\rm loc}}+ E{_u^{\rm trans}}\right) - \sum_{u=1}^{U}E{_u^{\rm es}} - \eta\left(E^{\rm fly}+E^{\rm hov}\right)$.

The actor-critic framework of the DDPG-based RMTO-Sub algorithm comprises an environment entity, actor network, critic network and replay buffer, as depicted in Fig. 3. In the following, we will introduce the DDPG-based RMTO-Sub algorithm processes in details.
\begin{figure}[t]
  \begin{center}
  \includegraphics[width=3.2in]{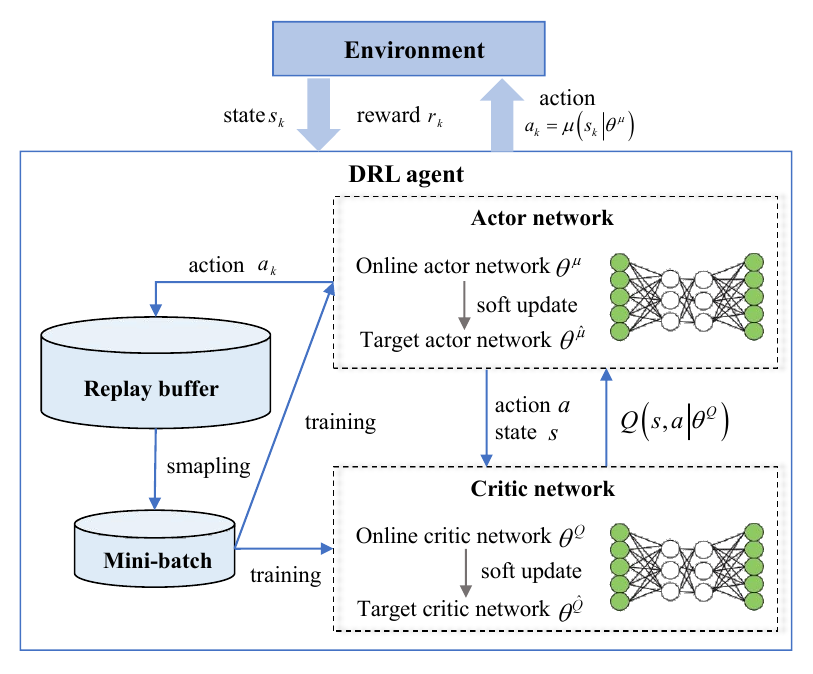}\\
  \caption{\textbf{Actor-critic framework of the DDPG-based RMTO-Sub algorithm.}}\label{circuit_diagram}
  \end{center}
\end{figure}

In the $k$th-round update, the action $a_k$ is chosen by the online actor network based on the current policy $\mu$, as follows:
\begin{equation}
	\begin{aligned}
		a_k=\mu(s_k|\theta^\mu),
	\end{aligned}
\end{equation}
where $\theta^\mu$ denotes the weight of online actor network. After selecting the action, the state is updated to $s_{k+1}$ and the agent receives an immediate reward denoted as $r_k$ through the interaction with the environment. Based on the Bellman equation, the action-value function by the online critic network with the weight $\theta^Q$ can be expressed as
\begin{equation}
	\begin{aligned}
		Q(s_k,a_k|\theta^Q)=\mathbb{E}_{\rho^\mu}[r_k(s_k,a_k)+\zeta Q(s_{k+1}, a_{k+1}|\theta^Q)],
	\end{aligned}
\end{equation}
where $\zeta \in \lbrack0,1)$ denotes the discounting factor, $\rho^\mu$ denotes state distribution of policy $\mu(s_k|\theta^\mu)$.
\begin{algorithm}[t]
        \caption{\textbf{DDPG-based RMTO-Sub Algorithm}}
        \LinesNumbered
		\KwIn{Maximum number of episodes $T$, maximum steps in each episode $K$, experience replay memory $\mathbb{R}$, mini-batch $\mathbb{I}, \mathbb{I}\subseteq\mathbb{R}$, learning rate of actor network $\alpha_\mu$, learning rate of critic network $\alpha_Q$, reward discount factor $\zeta$, soft update coefficient $\delta$.}
		\KwOut{Flying subslot $\boldsymbol{\Delta}^{\rm fly}$, hovering subslot $\boldsymbol{\Delta}^{\rm hov}$, offloading ratios of USVs $\boldsymbol{\alpha}$, computation capability allocation $\mathbf{F}$, UIRS trajectory $\mathbf{L}$.}
  
		{Initialize the weight of online actor network $\theta^\mu$ and the weight of online critic network $\theta^Q$.}
		
        {Initialize the weight of target actor network $\theta^{\hat{\mu}}$ and the weight of target critic network $\theta^{\hat{Q}}$.}
        
		{Initialize experience replay memory $\mathbb{R}$.}
  
        \For{episode $t \in \{1,2,...,T\}$}{
        
		{Initialize the state $s_1$.}
  
		\For {step $k \in \{1,2,...,K\}$}{
  
		{Update the state $s_k=\{\mathbf{l}_{{\rm uirs},k}, \mathbf{B}_k\}$.}
  
		{Select the action $a_k=\mu(s_k|\theta^\mu)$.}
  
		{Obtain reward $r_k$ and observe the next state $s_{k+1}$.}
  
		{Store experience $(s_k,a_k,r_k,s_{k+1})$ into the experience replay memory $\mathbb{R}$.}
  
		{Sample mini-batch $\mathbb{I}$ from the experience replay memory $\mathbb{R}$.}
  
		{Obtain the target function $y_k$ by (29).}
  
		{Update the online critic network by (30) and (31).}
  
		{Update the online actor network by (32) and (33).}
  
		{Update $\theta^{\hat{\mu}}$ and $\theta^{\hat{Q}}$ by (34) and (35), respectively.}
  
        }
        
		}
		\textbf{Return} $\boldsymbol{\Delta}^{\rm fly}, \boldsymbol{\Delta}^{\rm hov}, \boldsymbol{\alpha}, \mathbf{F}, \mathbf{L}$.
\end{algorithm} 

The target function can be calculated as 
\begin{equation}
	\begin{aligned}
		y_k=r_k(s_k,a_k)+\zeta\hat{Q}(s_{k+1},\hat{\mu}(s_{k+1}|\theta^{\hat{\mu}})|\theta^{\hat{Q}}),
	\end{aligned}
\end{equation}
where $\theta^{\hat{\mu}}$ and $\theta^{\hat{Q}}$ denote the weights of the target actor network and the target critic network, respectively. Online critic network is updated by minimizing the loss function as
\begin{equation}
	\begin{aligned}
		L(\theta^Q)=\mathbb{E}_{\rho^\mu}[(Q(s_k,a_k|\theta^Q)-y_k)^2].
	\end{aligned}
\end{equation}

The weight $\theta^Q$ updating of the online critic network is denoted by
\begin{equation}
	\begin{aligned}
		\theta^Q=\theta^Q+\alpha_Q\bigtriangledown_{\theta^Q}L(\theta^Q),
	\end{aligned}
\end{equation}
where $\alpha_Q$ denotes the learning rate of the critic network. With the sampling transitions from the experience replay memory, the online actor network with the weight $\theta^{\mu}$ is updated by the policy gradient method, as follows:
\begin{equation}
	\begin{aligned}
		&\bigtriangledown_{\theta^\mu}J \\
        &\quad =\mathbb{E}_{\rho^\mu}[\bigtriangledown_aQ(s_k,a_k|\theta^Q)|_{a=\mu(s_k|\theta^\mu)}\bigtriangledown_{\theta^\mu}\mu(s_k|\theta^\mu)].
	\end{aligned}
\end{equation}

The weight $\theta^\mu$ updating of the online actor network is denoted by
\begin{equation}
	\begin{aligned}
		\theta^\mu=\theta^\mu+\alpha_\mu\bigtriangledown_{\theta^\mu}J,
	\end{aligned}
\end{equation}
where $\alpha_\mu$ denotes the learning rate of the actor network. The weights of target actor network and target critic network are respectively updated as
\begin{equation}
	\theta^{\hat{\mu}}\leftarrow\delta\theta^\mu+(1-\delta)\theta^{\hat{\mu}},
\end{equation}
\begin{equation}
	\theta^{\hat{Q}}\leftarrow\delta\theta^Q+(1-\delta)\theta^{\hat{Q}},
\end{equation}
where $\delta$ denotes the soft update coefficient. The details of the DDPG-based RMTO-Sub algorithm is summarized in \textbf{Algorithm 3}.

The computational complexity of the DDPG-based RMTO-Sub algorithm depends on the actor and critic networks. The actor network has $L_a$ layers with $F_i$ neurons in the $i$th layer ($i \leq L_a$). The complexity of the $i$th layer is $\mathcal{O}(F_{i-1}F_i+F_iF_{i+1})$. The complexity of the actor network is $\mathcal{O}(\sum_{i=2}^{L_a-1}(F_{i-1}F_i+F_iF_{i+1}))$\upcite{r48}. The critic network has $L_c$ layers with $G_j$ in the $j$th layer ($j \leq L_c$). The complexity of the $j$th layer is $\mathcal{O}(G_{j-1}G_j+G_jG_{j+1})$. The complexity of the critic network is $\mathcal{O}(\sum_{j=2}^{L_c-1}(G_{j-1}G_j+G_jG_{j+1}))$. Thus, the complexity of the learning process for the DDPG-based RMTO-Sub algorithm is denoted as $\mathcal{O}(TK(\sum_{i=2}^{L_a-1}(F_{i-1}F_i+F_iF_{i+1})+\sum_{j=2}^{L_c-1}(G_{j-1}G_j+G_jG_{j+1})))$. 

\section{Numerical Results}
\subsection{Parameter Settings}
In this subsection, we conduct a set of simulations to demonstrate the effectiveness and efficiency of the proposed RMTO scheme. Assume that there are $U=4$ USVs randomly distributed in a target area of $1000$ m$\times$ $1000$ m. Referring to the parameter settings in Ref. \cite{r10,r28,r29,r30,r31,r32,r33,r34,r57,r58,r63}, the simulation parameters are shown as Table II. For the DDPG-based RMTO-Sub algorithm, the number of episodes $T$ is set to $1500$. The soft update coefficient $\delta$ and the reward discount factor $\zeta$ are set to $0.01$ and $0.9$, respectively. The learning rate of online actor network $\alpha_{\mu}$ and online critic network $\alpha_Q$ are set to $0.001$ and $0.001$, respectively. In addition, both the actor network and the critic network have three hidden layers with $(300,200,100)$ neurons.

\subsection{Performance Evaluation}
\begin{figure}[t]
    \centerline{\includegraphics[width=3.5in]{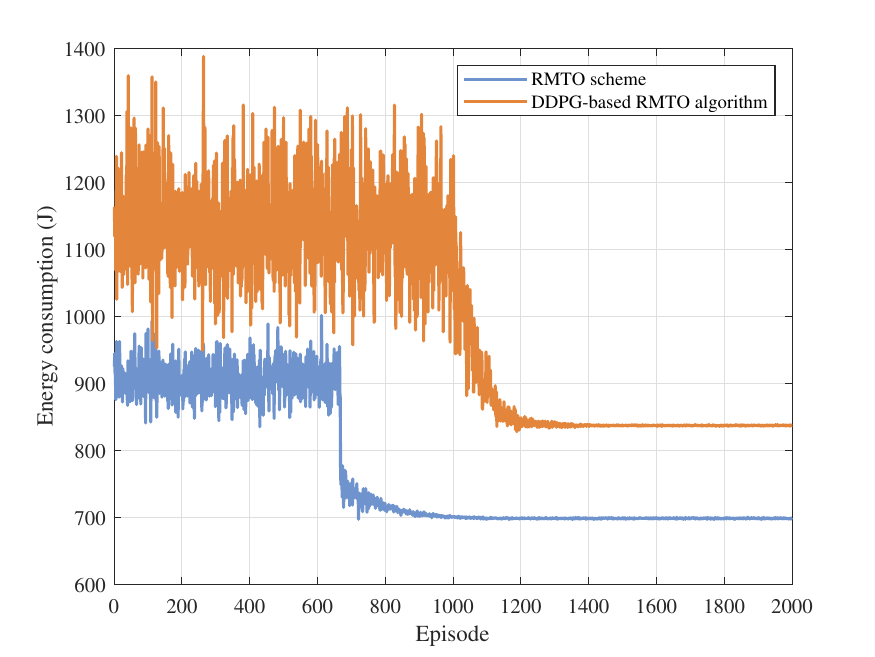}}
	\caption{\textbf{The convergence comparison of the proposed scheme and DDPG-based RMTO algorithm.}}
	\label{fig4}
\end{figure}

We first compare the convergence performance of the proposed RMTO scheme with the DDPG-based RMTO algorithm. ‘DDPG-based RMTO algorithm’ refers to the DDPG algorithm to solve the optimization problem in our work. Fig.~\ref{fig4} shows that the proposed scheme converges at $900$ episodes, while the DDPG-based RMTO algorithm converges after $1300$ episodes. The reason is that the complex state space and action space make the agent more challenging to explore. Therefore, the proposed scheme significantly improves the convergent speed and reduces the running time.
\begin{table}
\renewcommand{\arraystretch}{1.0}
	\caption{\textbf{Simulation Parameters}}
	\begin{center}
		\begin{tabular}{p{5.6cm} c}
			\hline
			\textbf{Parameter} & \textbf{Value}\\
			\hline
            The number of USVs $U$ & $4 \sim 10$ \\
			The number of time slots $K$ & $60$ \\
			The time span $\tau$ & $300$ s \\
			Maximum horizontal flying speed $v_{\rm max}^{\rm hor}$ & $30$ m/s \\
            Maximum vertical flying speed $v_{\rm max}^{\rm ver}$ & $8$ m/s \\
			Hovering power of UIRS $h_{\rm hov}$ & $43$ dBm \\
			Reference channel gain $\rho_0$ & $-80$ dB \\
			Channel bandwidth $W$ & $1$ MHz \\
			Gaussian noise power $\sigma^2$ & $-114$ dBm \\
			  Wavelength $\lambda$ & $0.05$ m\\
			Antenna height of USV $H_u$ & $1.9$ m \\
			Antenna height of mBS $H_{\rm mBS}$ & $5.45$ m \\
			Maximum transmit power of USV $P_u^{\rm max}$ & $44.7$ dBm \\
			Computation tasks $b_{u,k}$ & $[1,10]$ Mbit \\
			Required CPU cycles per bit at USV $C_u^{\rm loc}$ & $800$ cycles/bit \\
			Required CPU cycles per bit at edge server $C_{\rm es}$ & $1000$ cycles/bit \\
            Computation capability of USV $F_u$ & $0.1$ $\sim$ $0.2$ GHz \\
			Computation capability of edge server $F_{\rm es}^{\rm max}$ & $4$ GHz \\
			Switched coefficient for USV $\kappa_u^{\rm loc}$ & $10^{-27}$ \\
			Switched coefficient for edge server $\kappa_{\rm es}$ & $10^{-28}$ \\
			Tip velocity of the rotor blade $U_{\rm tip}$ & $120$ m/s \\
			Air density $\rho$ & $1.225$ kg/m$^3$ \\
			Rotor disc area $A_0$ & $0.503$ m$^2$ \\
			Rotor induced velocity in hovering $v^{\prime}$ & $4.03$ m/s \\
			Rotor solidity $S_0$ & $0.05$ \\
			Fuselage drag ratio $\xi$ & $0.6$ \\
			\hline
		\end{tabular}
		\label{tab1}
	\end{center}
\end{table}

\begin{figure}[t]
    \centering
    \subfigure[3D trajectory]{
        \includegraphics[width=3.2in]{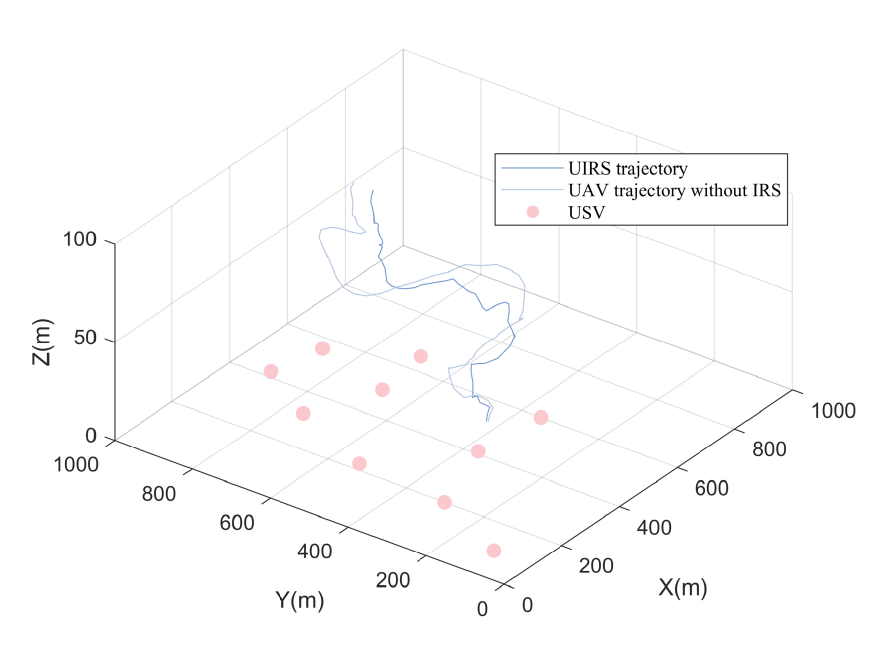}
    }
    \subfigure[2D trajectory]{
        \includegraphics[width=3.2in]{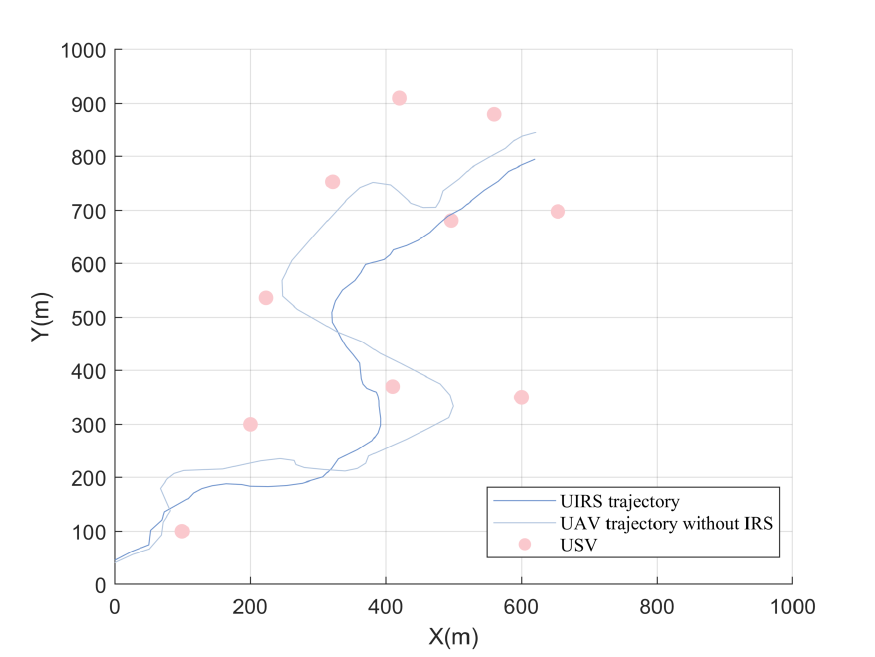}
    }
    \caption{\textbf{Illustrations of the UIRS trajectory and the UAV trajectory.}}
    \label{fig4b}
\end{figure}

Fig.~\ref{fig4b} (a) and (b) show the 2D and 3D trajectory of the UIRS and the UAV, respectively. Here, the UIRS trajectory is optimized by our proposed RMTO scheme, and the UAV trajectory means that the UAV acts as an amplify-and-forward relay node to achieve data transmission without the assistance of IRS. The two curves represent the trajectories, and the circles represent the locations of USVs. In the UAV trajectory without IRS scheme, the UAV flies close to the USV and drives to a lower altitude. This is because that the UAV approaches each USV so as to establish communication links to USVs for computation offloading, at the expense of more energy consumption. In contrast, in the RMTO scheme, the UIRS flight path is much shorter. This is due to the fact that the UIRS can balance the cascaded channel gain between the USVs and the mBS to provide computation offloading service, with the help of the flexible UIRS in the RMTO scheme.

To demonstrate the efficiency of the RMTO scheme, we compare the performance of the RMTO scheme with five baseline schemes.

$\bullet$ \textit{Fixed Computation Capability Allocation (FCCA) scheme}: The RMTO scheme is applied without computation capability allocation optimization. The edge computation capability is equally allocated to each USV for processing computation tasks.

$\bullet$ \textit{Frequency Division Multiple Access (FDMA) scheme}: The RMTO scheme adopts FDMA instead of NOMA during transmission. The specific bandwidth is equally allocated for computation offloading from USVs to the UIRS. 

$\bullet$ \textit{UIRS straight flying scheme}: The RMTO scheme is applied without UIRS trajectory optimization. The UIRS flies from the initial position to the final position following a straight trajectory.

$\bullet$ \textit{UIRS hovering scheme}: The RMTO scheme is applied without UIRS trajectory optimization. The UIRS is hovering at the fixed position to assist transmission from USVs to the edge server.

$\bullet$ \textit{Full offloading scheme}: The RMTO scheme is applied without offloading ratio optimization. Computation task data is uniformly offloaded to the edge server for edge computing.

\begin{figure}[t]
	\centerline{\includegraphics[width=3.5in]{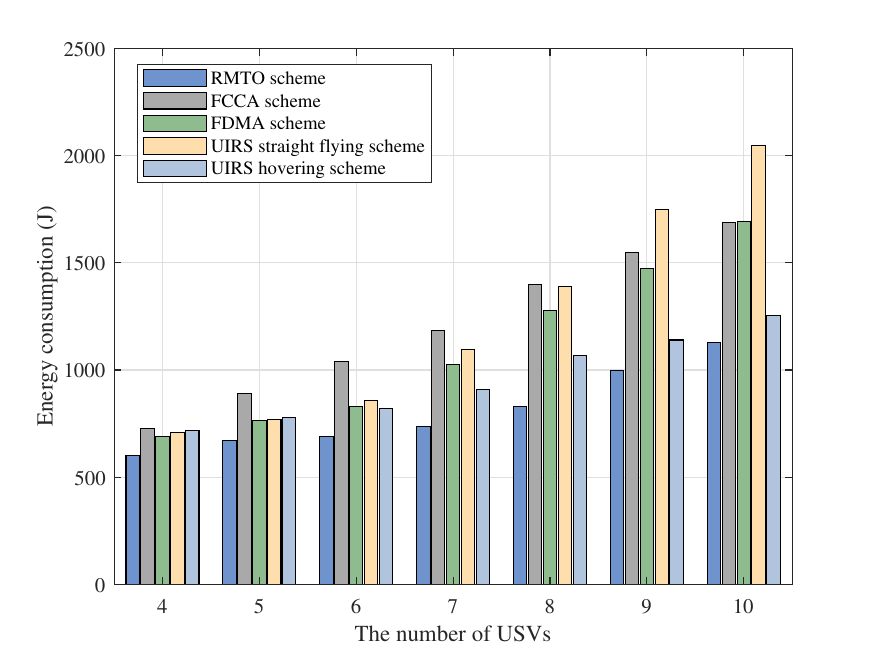}}
	\caption{\textbf{Energy consumption varies with the number of USVs.}}
	\label{fig5}
\end{figure}

Fig.~\ref{fig5} illustrates the overall energy consumption varies with the number of USVs. The result shows that the RMTO scheme exhibits a significant advantage over the FCCA, FDMA, UIRS straight flying and UIRS hovering schemes, achieving an average of $24.4\%$, $14.9\%$, $31.8\%$ and $29.2\%$ reduction in the overall energy consumption, respectively. According to the comparison between without UIRS trajectory optimization scheme (e.g., UIRS straight flying scheme and UIRS hovering scheme) and the RMTO scheme, it can be concluded that the resource management will be able to show its huge performance advantages if the UIRS trajectory optimization is carried out. These phenomena demonstrate that the proposed joint resource management and the UIRS trajectory optimization can reduce overall energy consumption.

\begin{figure}[t]
	\centerline{\includegraphics[width=3.5in]{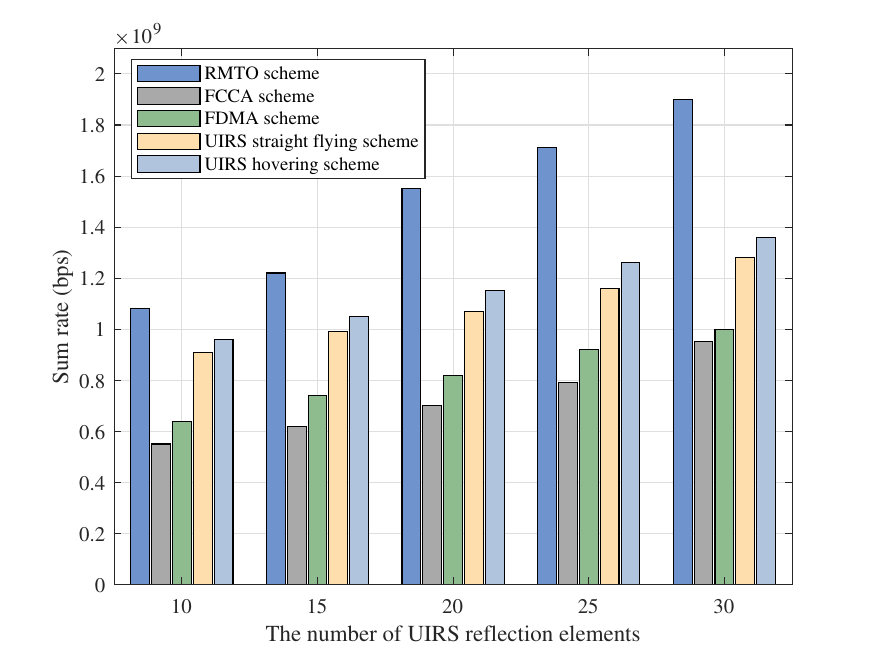}}
	\caption{\textbf{Achievable sum rate varies with the number of UIRS reflection elements.}}
	\label{fig6}
\end{figure}

\begin{figure}[t]
	\centerline{\includegraphics[width=3.5in]{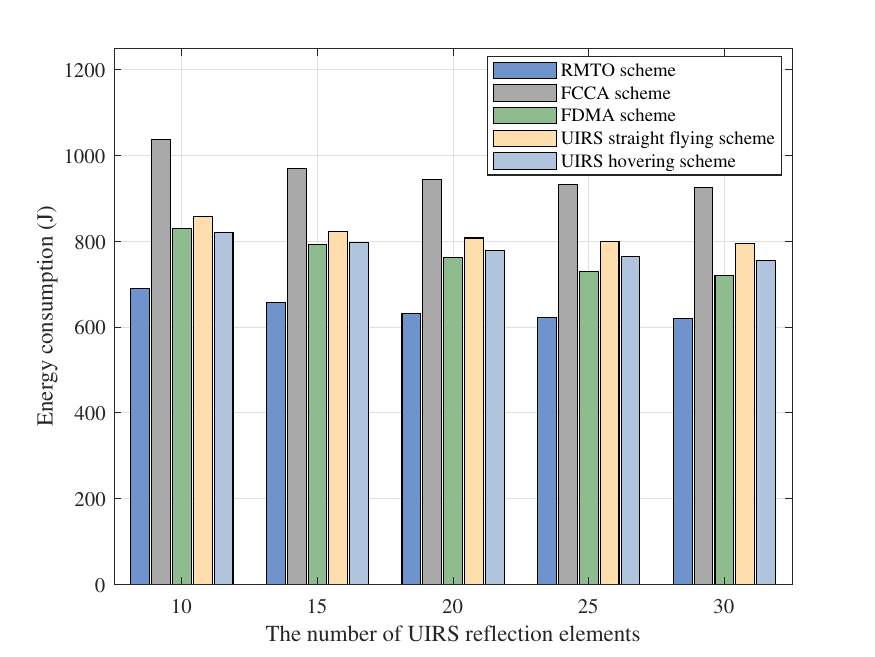}}
	\caption{\textbf{Energy consumption varies with the number of UIRS reflection elements.}}
	\label{fig7}
\end{figure}

Fig.~\ref{fig6} and Fig.~\ref{fig7} illustrate the trend of sum rate and energy consumption as the number of UIRS reflection elements changes based on five schemes, respectively. Here, the sum rate is defined as the cumulative data rate of all the USVs. It can be seen that the sum rate increases and the energy consumption reduces, as the number of UIRS reflection elements increases. The reason is that with the increased number of UIRS reflection elements enables the beamforming gain to improve, thus increasing the sum rate of USVs. However, when expanding from $25$ to $30$ UIRS reflection elements, the energy consumption only has a $0.55\%$ reduction. Therefore, a tradeoff between the UIRS reflection elements and the number of USVs needs to be considered.

\begin{figure}[t]
	\centerline{\includegraphics[width=3.5in]{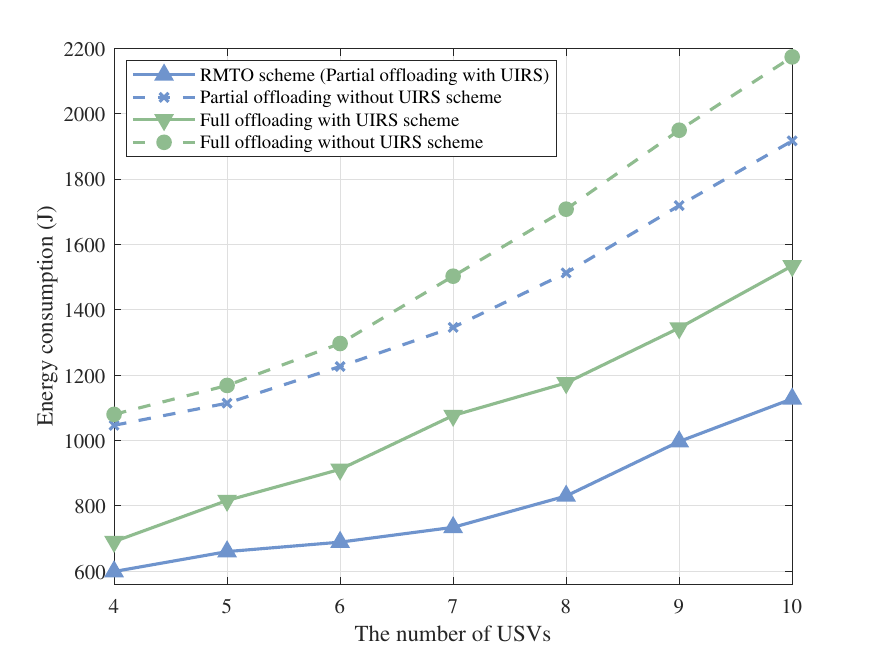}}
	\caption{\textbf{Comparison energy consumption between partial offloading and full offloading schemes in different numbers of USVs.}}
	\label{fig8}
\end{figure}

Fig.~\ref{fig8} shows the effectiveness of UIRS and partial offloading employed in our work. For the scheme without UIRS, it means that the UAV acts as an amplify-and-forward relay node to assist task data transmission from USVs to the mBS. It can be seen that employing UIRS reduces the overall energy consumption compared with the scheme without UIRS, and the energy consumption saved by UIRS is more prominent when the number of USVs increases. Moreover, compared to the full offloading scheme, the RMTO scheme greatly reduces energy consumption, and the significant energy savings increase as the number of USVs increases. 

\begin{figure}[t]
	\centerline{\includegraphics[width=3.5in]{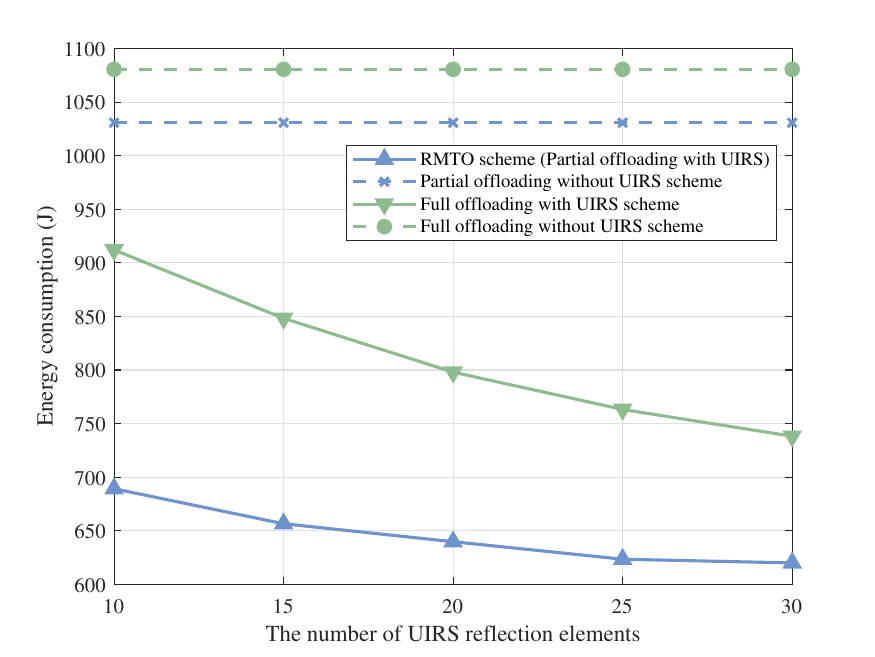}}
	\caption{\textbf{Comparison energy consumption between partial offloading and full offloading schemes in different numbers of UIRS reflection elements.}}
	\label{fig9}
\end{figure}

Fig.~\ref{fig9} shows the relationship between the energy consumption and the number of UIRS reflection elements based on partial offloading scheme and full offloading scheme. As the number of UIRS reflection elements increases, the higher data rate can be obtained, which can transmit more data bits to the mBS with low transmit power, decreasing the transmission energy consumption and local computation energy consumption. Besides, the energy consumption gap between the RMTO scheme and the full offloading with UIRS scheme decreases as the number of UIRS reflection elements increases. The results in Fig.~\ref{fig8} and Fig.~\ref{fig9} imply the effectiveness of the proposed scheme in optimizing offloading ratios of USVs and employing UIRS for UIRS-assisted maritime edge computing networks.

\begin{figure}[t]
	\centerline{\includegraphics[width=3.5in]{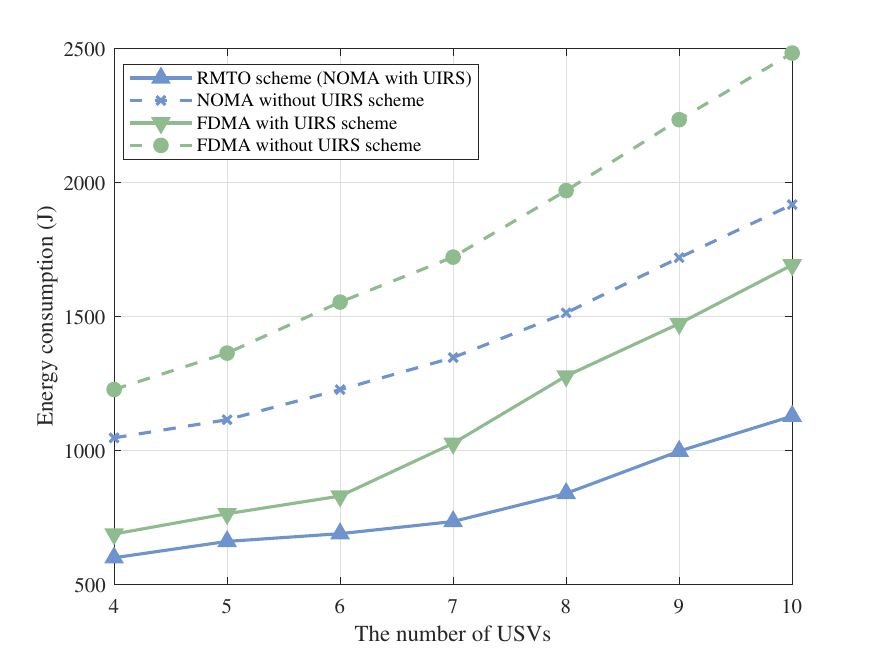}}
	\caption{\textbf{Comparison energy consumption between NOMA and FDMA schemes in different numbers of USVs.}}
	\label{fig10}
\end{figure}

Fig.~\ref{fig10} shows the efficiency of the RMTO scheme in adopting the NOMA, in comparison with the FDMA scheme. The energy consumption for the FDMA with UIRS scheme is greater than that for the NOMA without UIRS scheme. It means the sacrifice of channel bandwidth in the FDMA can be compensated by employing the UIRS. Furthermore, the disparity in energy consumption between the RMTO scheme and FDMA with UIRS scheme escalates as the number of USVs increases, which indicates that the NOMA performs better than the FDMA scheme, especially in large-scale users. Therefore, the results imply the effectiveness of the RMTO scheme in adopting NOMA and employing UIRS for maritime edge computing networks.

\begin{figure}[t]
	\centerline{\includegraphics[width=3.5in]{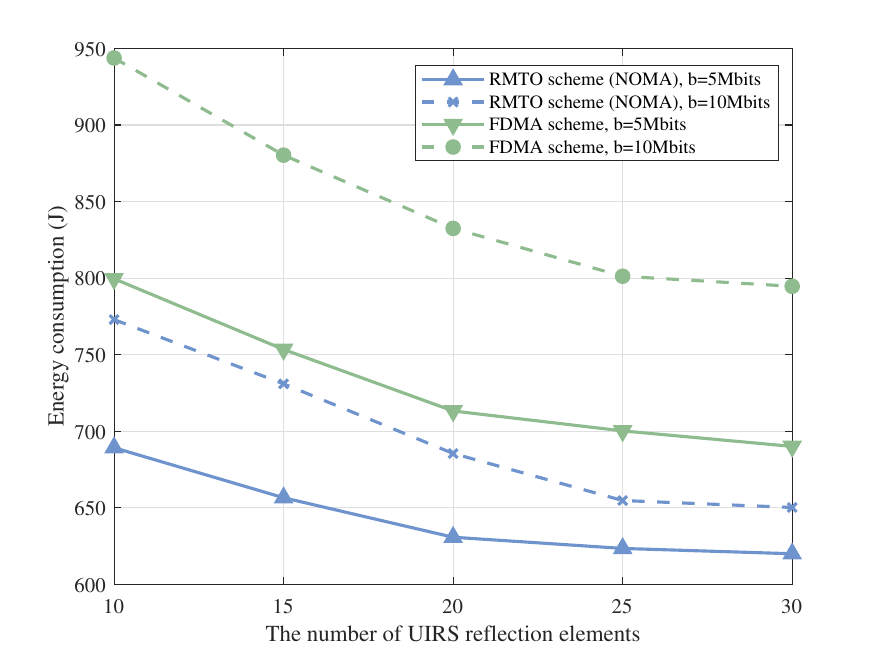}}
	\caption{\textbf{Comparison energy consumption between NOMA and FDMA schemes in different numbers of UIRS reflection elements.}}
	\label{fig11}
\end{figure}

Fig.~\ref{fig11} shows the relationship between the energy consumption and the number of UIRS reflection elements based on NOMA scheme and FDMA scheme. It can be seen that the energy consumption decreases with the number of UIRS reflection elements increases. Specifically, the disparity in energy consumption between $b=5$ Mbits scheme and $b=10$ Mbits scheme decreases as the number of UIRS reflection elements increases. It implies that even large amount of task data to process, the energy consumption can be reduced through the increasing number of UIRS reflection elements. The results in Fig.~\ref{fig10} and Fig.~\ref{fig11} imply the effectiveness of the proposed scheme in adopting NOMA and employing UIRS for maritime edge computing networks.

\section{Conclusion}
This paper has proposed resource management and trajectory optimization scheme for UIRS-assisted maritime edge computing networks. The objective is to minimize the overall energy consumption via jointly optimizing subslot duration, offloading ratios of USVs, transmit power of USVs, computation capability allocation, UIRS phase shifts and UIRS trajectory. To solve the non-convex optimization problem, we have proposed a two-layered method by decomposing the original problem into two subproblems to attain the near optimal solution. In the RMTO-Top optimization stage, we have derived the closed-form expression of transmit power of USVs and employed the SDR method to optimize UIRS phase shifts. In the RMTO-Sub optimization stage, we have proposed a DDPG-based RMTO-Sub algorithm to optimize subslot duration, offloading ratios of USVs, computation capability allocation and UIRS trajectory. Numerical results have demonstrated the convergence performance of our proposed RMTO scheme. It has also been discovered that the RMTO scheme achieves a significant performance improvement in different simulation environments in comparison with baseline schemes. 

We have investigated the performance of integration of communication and computation for UIRS-assisted maritime edge computing networks. In future work, it is possible that IRS can be utilized as a sensing facility to study the network performance by designing the passive beamforming in the integration of sensing, communication and computation networks. Moreover, an open problem is to study the secrecy rate enhancement under the eavesdropping attack in the UIRS-assisted physical layer security.  

\vskip 2mm
\large
\noindent
\textbf{Acknowledgment}
\vskip 2mm

\Acknow
\noindent
The work was supported in part by the National Natural Science Foundation of China under Grants 62371085, 51939001, 62122069, and 62071431, and in part by the Fundamental Research Funds for the Central Universities under Grant 3132023514.

\renewcommand\refname{\large\textbf{References}}
  
\bibliographystyle{IEEEtrans}   
\bibliography{REF}


\begin{biography}[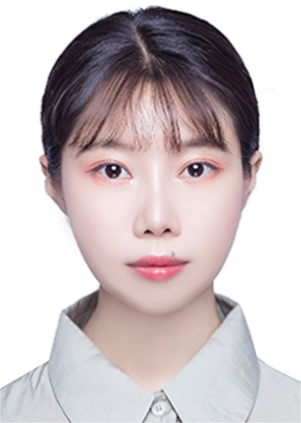]
 \noindent
 \textbf{Chaoyue Zhang}\ \  received the B.S. degree from Wuhan University of Science and Technology, Wuhan, China, in 2017, and the M.S. degree from Dalian Maritime University, Dalian, China, in 2020, where she is currently pursuing the Ph.D. degree with the Information Science and Technology College. Her research interests include maritime communication networks, mobile edge computing, and resource management.
\end{biography}

\begin{biography}[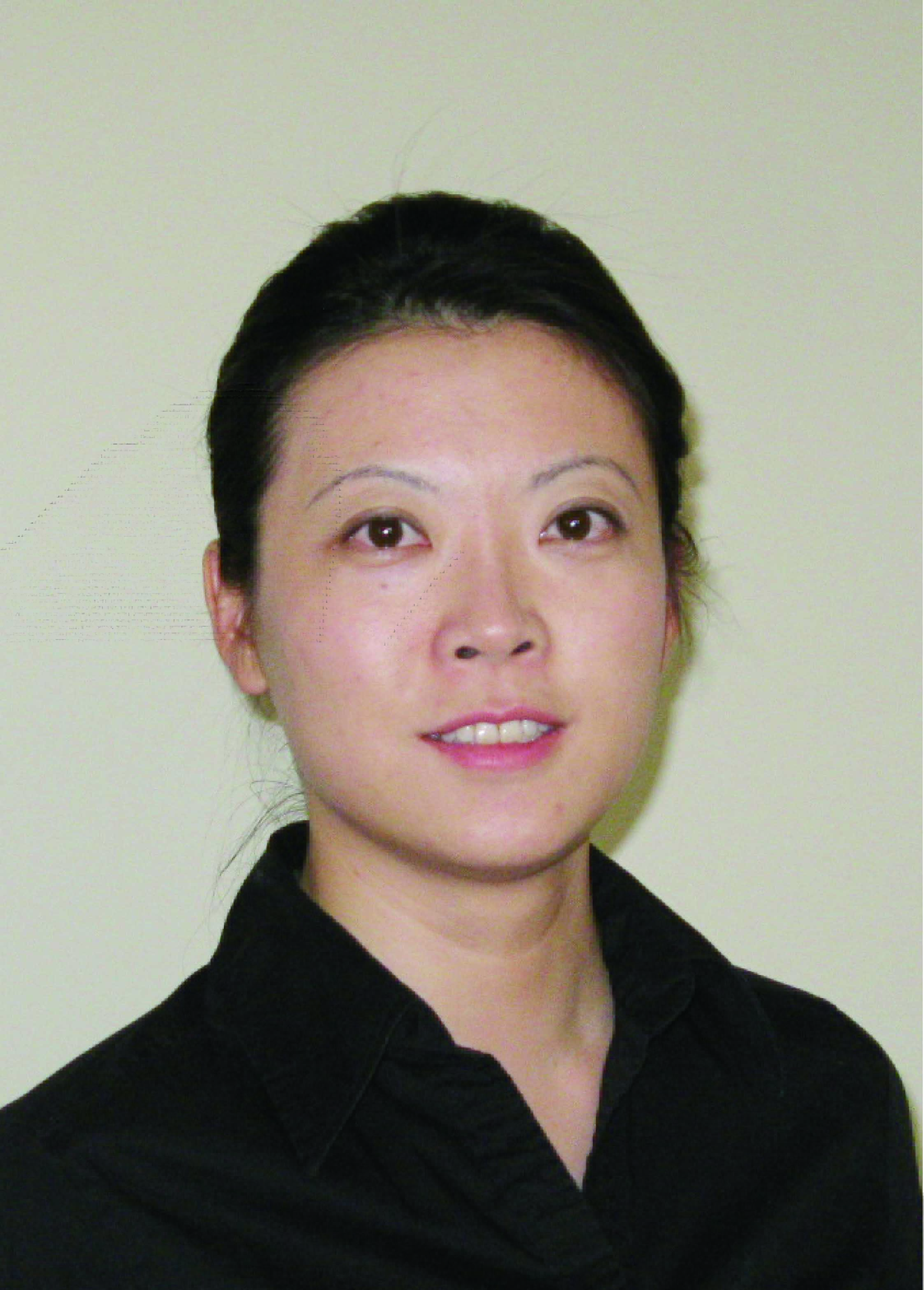]
\noindent
\textbf{Bin Lin}\ \ received the B.S. and M.S. degrees from Dalian Maritime University, Dalian, China, in 1999 and 2003, respectively and the Ph.D. degree from the Broadband Communications Research Group, Department of Electrical and Computer Engineering, University of Waterloo, Waterloo, ON, Canada, in 2009. She is currently a Full Professor and the Dean of Communication Engineering with the College of Information Science and Technology, Dalian Maritime University. She has been a Visiting Scholar with George Washington University, Washington, DC, USA, from 2015 to 2016. Her current research interests include wireless communications, network dimensioning and optimization, resource allocation, artificial intelligence, maritime communication networks, and Internet of Things. Prof. Lin is an associate editor of IEEE Transactions on Vehicular Technology and IEEE Internet of Things Journal.
\end{biography}

\begin{biography}[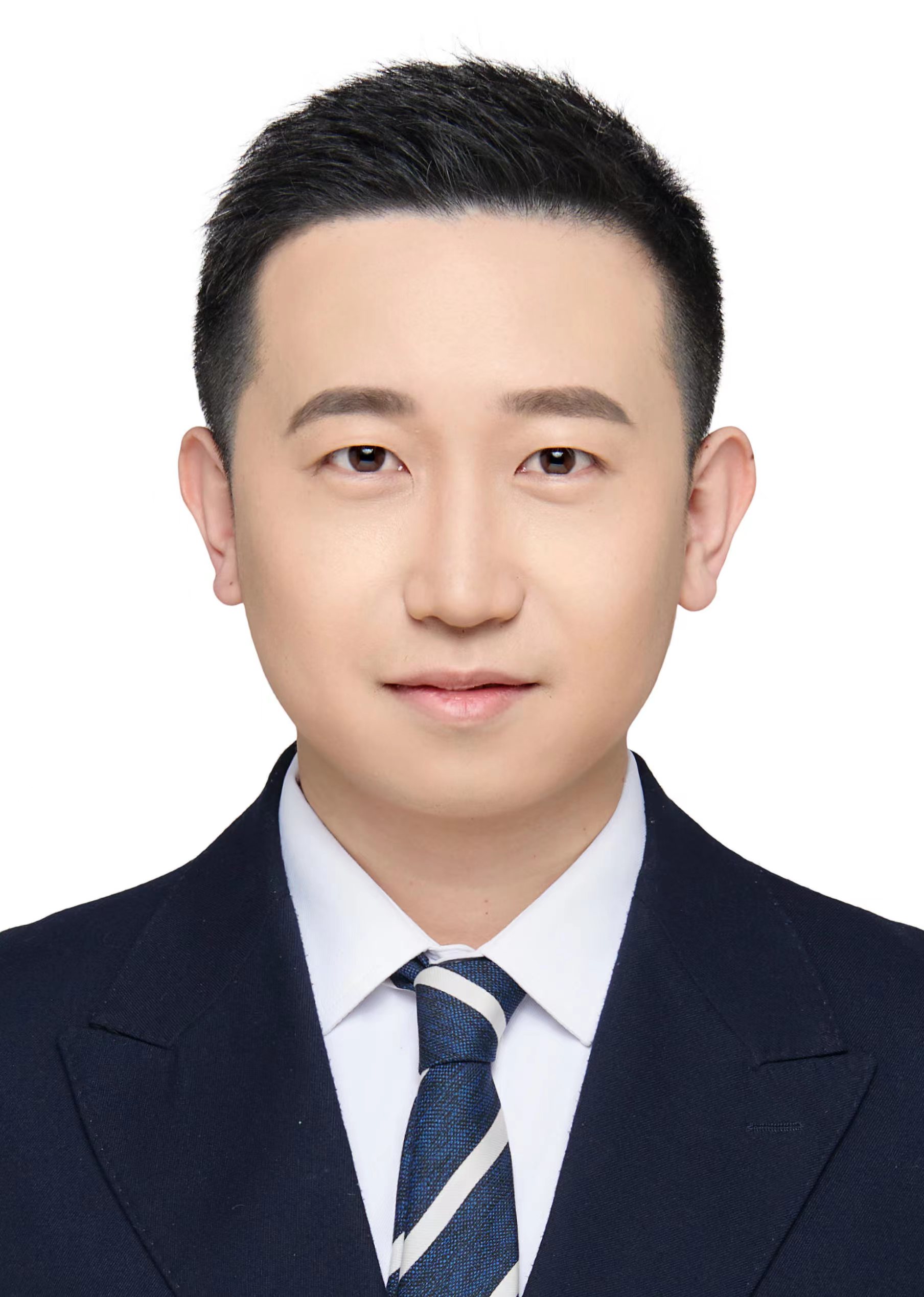]
\noindent
\textbf{Xu Hu}\ \  received the B.S. degree from Dalian Polytechnic University, Dalian, China, in 2017, and the M.S. degree from Dalian Maritime University, Dalian, in 2020, where he is currently pursuing the Ph.D. degree with the Information Science and Technology College. His research interests include maritime communication networks, space-air-ground-sea integrated communication networks, and stochastic geometry.
\end{biography}

\begin{biography}[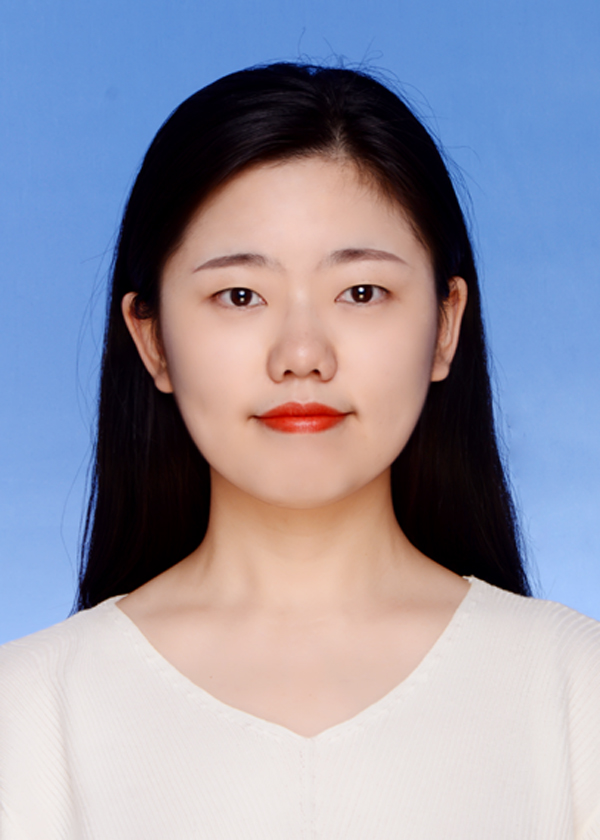]
\noindent
\textbf{Shuang Qi}\ \ received the B.S. and M.S. degrees from Liaoning Normal University, Dalian, China, in 2017 and 2020, respectively. She is currently pursuing the Ph.D. degree with the College of Information Science and Technology, Dalian Maritime University, Dalian, China. Her research interests include maritime communication networks, mobile edge computing, and resource management.
\end{biography}

\begin{biography}[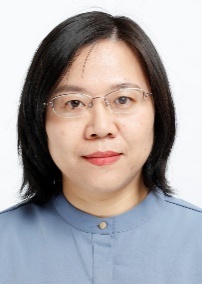]
\noindent
\textbf{Liping Qian} \ \ received the PhD degree in Information Engineering from the Chinese University of Hong Kong in 2010. She worked as a postdoctoral research associate at the Chinese University of Hong Kong, China, during 2010-2011. Since 2011, she has been with College of Information Engineering, Zhejiang University of Technology, Hangzhou, China, where she is currently a Full Professor. From 2016 to 2017, she was a visiting scholar with the Broadband Communications Research Group, ECE Department, University of Waterloo. Her research interests include wireless communication and networking, resource management in wireless networks, massive IoTs, mobile edge computing, emerging multiple access techniques, and machine learning oriented towards wireless communications. She was a co-recipient of the IEEE Marconi Prize Paper Award in Wireless Communications in 2011, the Best Paper Award from IEEE ICC 2016, the Best Paper Award from IEEE Communication Society GCCTC 2017, the Best Paper Award from the Digital Communications and Networking in 2021, and the Best Paper Award from IEEE WCNC 2023. She was an Associate Editor of the IET Communications from 2016 to 2022. She is currently on the Editorial Boards of IEEE Transactions on Cognitive Communications and Networking and IEEE Internet of Things Journal.
\end{biography}

\begin{biography}[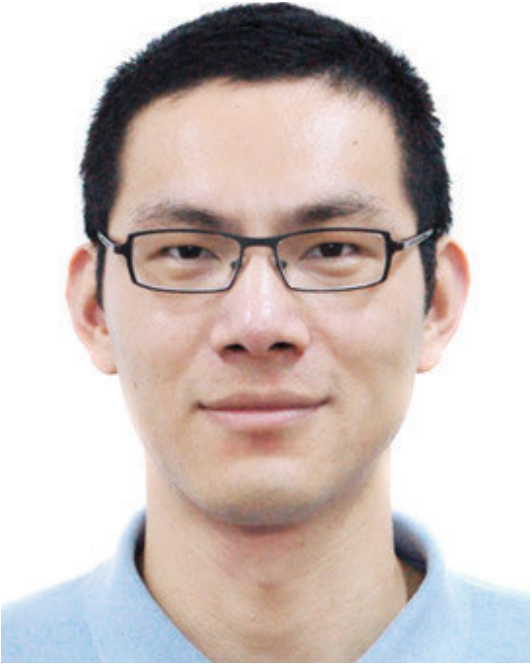]
\noindent
\textbf{Yuan Wu} \ \ is currently an Associate Professor with the State Key Laboratory of Internet of Things for Smart City, University of Macau, Macao, China, and also with the Department of Computer and Information Science, University of Macau. He received the PhD degree in Electronic and Computer Engineering from the Hong Kong University of Science and Technology in 2010. His research interests include resource management for wireless networks, green communications and computing, edge computing and edge intelligence, and energy informatics. He received the Best Paper Award from the IEEE ICC'2016, IEEE TCGCC'2017, IWCMC'2021, and IEEE WCNC'2023. Dr. Wu is currently on the editorial board of IEEE Transactions on Vehicular Technology, IEEE Transactions on Network Science and Engineering, and IEEE Internet of Things Journal. 
\end{biography}


\mbox{}
\clearpage
\clearpage
\large
\noindent

\end{document}